\documentclass[preprint,12pt]{elsarticle}

\usepackage{amssymb}
\usepackage{amsmath}
\usepackage{amsthm}

\usepackage[switch]{lineno}
\usepackage{hyperref}
\usepackage{multirow}
\usepackage{subcaption}
\journal{Icarus}
\usepackage{url}
\usepackage{longtable}
\usepackage{float}
\usepackage{pdflscape}

\usepackage{placeins}

\usepackage{xcolor}
\usepackage[normalem]{ulem}

\begin{document}

\begin{frontmatter}



\title{Cross-Calibration of Hera/HyperScout-H with MRO/CRISM and MEx/OMEGA Based on Mars Observations}

\author[1]{A. Farina}
\author[2,3]{M. Popescu}
\author[1]{M. Lazzarin}
\author[4,5]{J. de León}
\author[4,5]{G. Prodan}
\author[2]{B. A. Dumitru}
\author[1]{F. La Forgia}
\author[6]{G. Poggiali}
\author[7]{M. Küppers}
\author[8]{G. Kovács}
\author[8]{B. V. Nagy}
\author[9]{B. Grieger}
\author[10]{O. Barnouin}
\author[4,5]{E. Tatsumi}
\author[11]{\"{O}. Karatekin}
\author[12]{P. Michel}

\affiliation[1]{organization={Dipartimento di Fisica e Astronomia, Padova University}, addressline={Vicolo dell'Osservatorio 3}, city={Padova}, postcode={35122}, country={Italy}}
\affiliation[2]{organization={Institute of Space Science – INFLPR Subsidiary}, addressline={409 Atomi\c{s}tilor Street}, city={M\u{a}gurele}, postcode={077125}, state={Ilfov}, country={Romania}}
\affiliation[3]{organization={University of Craiova}, addressline={Alexandru Ioan Cuza 13}, city={Craiova}, postcode={200585}, country={Romania}}
\affiliation[4]{organization={Instituto de Astrofísica de Canarias}, addressline={C/Vía Láctea s/n}, city={La Laguna}, postcode={E-38205}, state={Tenerife}, country={Spain}}
\affiliation[5]{organization={Department of Astrophysics, University of La Laguna}, addressline={C/Astrofísico Francisco Sánchez, s/n}, city={La Laguna}, postcode={E-38205}, state={Tenerife}, country={Spain}}
\affiliation[6]{organization={INAF-Osservatorio Astrofisico di Arcetri}, addressline={Largo Enrico Fermi, 5}, city={Firenze}, postcode={50125}, country={Italy}}
\affiliation[7]{organization={ESA, European Space Astronomy Centre (ESAC)}, city={Madrid}, country={Spain}}
\affiliation[8]{organization={Department of Mechatronics, Optics and Engineering Informatics, Budapest University of Technology and Economics}, city={Budapest}, country={Hungary}}
\affiliation[9]{organization={Aurora Technology B.V. for ESA, ESAC}, city={Madrid}, country={Spain}}
\affiliation[10]{organization={Space Exploration Sector, Johns Hopkins University Applied Physics Laboratory}, city={Laurel}, state={Maryland}, country={USA}}
\affiliation[11]{organization={Royal Belgian Institute for Space Aeronomy}, addressline={Av. Circulaire 3}, postcode={1180}, city={Brussels}, country={Belgium}}
\affiliation[12]{organization={Université Côte d'Azur, Observatoire de la Côte d'Azur, CNRS, Laboratoire Lagrange}, city={Nice}, country={France}}

\begin{abstract}

On March~12, 2025, Hera, the European Space Agency's planetary defense mission, performed a Mars flyby en route to the asteroid system (65803) Didymos--Dimorphos. During this event, HyperScout-H (HS-H), the spacecraft's hyperspectral imager, acquired five high spatial resolution images over the southern hemisphere of Mars. These observations provide an opportunity for systematic cross-calibration of HS-H with two established Mars orbital instruments, Compact Reconnaissance Imaging Spectrometer for Mars (CRISM) on board the Mars Reconnaissance Orbiter \citep{Murchie2007}, and Observatoire pour la Min\'eralogie, l'Eau, les Glaces et l'Activit\'e (OMEGA) on board Mars Express \citep{Bibring2004}.
This study evaluates the performance, reliability, and readiness of HS-H for the forthcoming asteroid observation phase by comparing it with CRISM and OMEGA data. We selected  21 surface regions spanning diverse terrain types from three HS-H images in which the Martian surface fills the entire field of view. These images were corrected, calibrated, and co-registered to associate each pixel with latitude and longitude.
The relative inter-channel correction factors are  close to unity for bands~3--25 (0.688--0.952~$\mu$m), with standard deviations of  0.52$\%$ for OMEGA and  0.62$\%$ for CRISM, indicating good agreement in spectral shape. Bands~1 and~2 show a systematic mismatch of  4.7$\%$ and 3.6$\%$. The absolute flux correction factors reveal offsets of  8$\%$ relative to OMEGA and approximately  0.7$\%$ relative to CRISM, with mean inter-image standard deviations below 2$\%$. These results confirm the radiometric stability of HS-H and establish a calibration framework for future Didymos--Dimorphos observations.

\end{abstract}



\begin{keyword}
Instrumentation \sep Asteroids \sep Spectrophotometry \sep Near-Earth objects \sep Mars


\end{keyword}

\end{frontmatter}




\section{Introduction}
\label{sec:intro}

The Hera mission was developed within the Space Safety Programme of the European Space Agency (ESA) as the European contribution to the Asteroid Impact and Deflection Assessment (AIDA) collaboration. Its target is the binary system Didymos–Dimorphos, whose secondary, Dimorphos, was the impact site of NASA's Double Asteroid Redirection Test (DART), the first demonstration of asteroid deflection by means of a kinetic impactor \citep{Rivkin2021, Chabot2024}. Building on this experiment, Hera will characterize in detail the outcome of the DART impact and determine the physical, surface, and internal properties of both components of the system, providing the ground truth needed to validate impact models for planetary defense \citep{Michel2018, Michel2022}. \\

To achieve these goals, Hera adopts a novel architecture for deep-space exploration, combining a main spacecraft with two CubeSats, Juventas and Milani \citep{Michel2022}. The main spacecraft carries the Asteroid Framing Cameras (AFC), the Hyperspectral Imager (HyperScout-H, hereafter HS-H), the Thermal Infrared Imager (TIRI), and the Planetary Altimeter (PALT). Juventas hosts the low-frequency radar JuRa, devoted to the first direct sounding of an asteroid's interior, and the gravimeter GRASS, while Milani carries the near-infrared imaging spectrometer ASPECT and the VISTA thermogravimeter for dust characterization. Hera was launched on 7 October 2024 at 14:52 UT from Cape Canaveral aboard a SpaceX Falcon 9 rocket \citep{Michel2025}. A few days after launch, it acquired images of the Earth–Moon system that confirmed the nominal operation of its onboard instruments, and on 12 March 2025 it performed a successful flyby of Mars, during which it also observed the Martian moon Deimos \citep{Michel2025}. Hera will reach the Didymos–Dimorphos system in early November 2026, when it will begin the proximity phase and observe the binary for at least six months. \\

Didymos is a binary near-Earth asteroid belonging to the Apollo dynamical family. The primary has a diameter of $761 \pm 26$~m, while the secondary, called Dimorphos, measures $151 \pm 5$~m \citep{Daly2023}. From a spectroscopic point of view, the system is classified as an S-type asteroid \citep{DeLeon2006}, linked to ordinary chondrites \citep{Dunn2013}. After the DART impact, the surface composition remains consistent with an S-type classification, with small variations in the spectral slope that may be related to the evolution of the ejecta cloud or to material deposited on the surface of Didymos \citep{Polishook2023, Lazzarin2026}. The orbital period of Dimorphos decreased by approximately 33~minutes as a result of the impact \citep{Scheirich2024, Naidu2024}. \\

The HS-H instrument is a hyperspectral imager onboard the Hera spacecraft, acquiring simultaneous spatial and spectral information over the 0.65–0.95 $\mu$m wavelength range. Its primary scientific objectives are to characterize the surface composition of Didymos and Dimorphos, investigate space-weathering effects, and search for exogenous material on these bodies \citep{Popescu2025}. In addition, HS-H will reveal the spectral properties of freshly exposed subsurface material excavated by the DART impact and produce spectral maps of taxonomic classification, spectral slope, and band parameters to better constrain the system's physical properties \citep{Michel2022, Popescu2025}. During the March 2025 Mars flyby, HS-H acquired observations of Mars and its two moons, Phobos and Deimos, providing a valuable opportunity to assess the instrument's in-flight radiometric performance before its arrival at the Didymos system in November 2026. \\

To validate the HS-H observations, we made use of data from two well-established hyperspectral instruments that have been observing Mars from orbit for over two decades, namely the \emph{Observatoire pour la Minéralogie, l'Eau, les Glaces et l'Activité} (OMEGA) onboard ESA's Mars Express spacecraft \citep{Bibring2004} and \emph{Compact Reconnaissance Imaging Spectrometer for Mars} (CRISM) onboard NASA's Mars Reconnaissance Orbiter \citep{Murchie2007}. Both instruments cover wavelength ranges that overlap with HS-H, with OMEGA spanning 0.35–5.2 $\mu$m and CRISM covering 0.362–3.92 $\mu$m, making them ideal spectral references in the HS-H wavelength range. Their long operational history and extensive calibration heritage provide a robust and well-characterized dataset against which the HS-H data can be meaningfully compared. \\

OMEGA is a hyperspectral imaging spectrometer onboard Mars Express, in orbit around Mars since December 2003 \citep{Bibring2004}. It covers the visible and near-infrared spectral range from 0.35 to 5.2~$\mu$m in 352 contiguous spectral  samples, with a spatial resolution ranging from 300~m to 4.8~km per pixel depending on the orbital altitude. The term channel is here reserved for the three spectrometers of OMEGA, namely VNIR, SWIR-C and SWIR-L. The VNIR channel, with 96 spectral samples, covers the HS-H spectral range and therefore provides the data set relevant to this comparison. The spectral samples beyond 0.93~\textmu m are affected by contamination from high grating orders and are not delivered in the data products, which leaves a small gap in the overlap with HS-H between 0.93 and 0.95~\textmu m. Its primary scientific objective is the global mineralogical mapping of the Martian surface, with particular focus on detecting and characterizing hydrated minerals, ices, and volcanic deposits \citep{Bibring2005}. \\

CRISM is a hyperspectral imaging spectrometer onboard the Mars Reconnaissance Orbiter (MRO), which has been orbiting Mars since 2006 \citep{Murchie2007}. It operates in the visible and near-infrared range from 0.362 to 3.92~$\mu$m with 544 spectral channels at a sampling of 6.55~nm per channel. In targeted mode, CRISM achieves a spatial resolution of 18 or 36~m~pixel$^{-1}$, significantly higher than OMEGA, allowing the characterization of small-scale mineralogical deposits and the identification of aqueous alteration products that cannot be resolved at OMEGA's spatial scale. In survey mode, the spatial resolution is 100 or 200~m~pixel$^{-1}$ with 72 spectral samples per pixel \citep{Murchie2007, Viviano2014}. The CRISM data used in this work are the VNIR hyperspectral mapping products described in Sect.~\ref{subsec:instruments}, which retain the full spectral resolution of the VNIR detector.\\

Cross-calibration between space instruments is a necessary step to assess the reliability and repeatability of scientific measurements from different missions. Independent instruments can show systematic offsets in their absolute radiometric calibration. These offsets may arise from differences in ground calibration procedures, adopted solar irradiance models, and photometric correction pipelines. If these offsets are not taken into account, spectra acquired by different instruments cannot be compared directly. This may lead to incorrect interpretations of the surface spectral properties. For HS-H, cross-calibration is therefore required to validate the instrument's in-flight performance. It also allows HS-H data to be compared with the spectral datasets acquired by previous Mars orbiters. This provides a consistent reference for the analysis of future observations of the Didymos-Dimorphos system. Beyond the instrument itself, science observations performed during the cruise phase provide an early validation, under real operating conditions, of the entire processing chain: from the raw data acquired by the instrument to spacecraft telemetry, the mission ground segment, the retrieval of Level 1A data and their conversion into radiometric quantities, and the determination of the observing conditions, namely distance, photometric angles, and projected instantaneous field of view on the target, from navigation data. An extended and well-characterized target such as Mars is particularly suited to this verification, which cannot be carried out on the unresolved objects observed during cruise. \\

In this work, we present a comprehensive cross-calibration of the HS-H instrument against OMEGA and CRISM over 21 surface regions of Mars, deriving wavelength-dependent correction factors for all 25 HS-H spectral bands in both normalized and absolute flux. This paper is organized as follows. Section~\ref{sec:DataObservations} describes the HS-H instrument and the datasets used. Section~\ref{sec:OMEGA_CRISM} describes the OMEGA 
and CRISM  data sets, the spectral extraction pipeline, and the spectral preprocessing. Section~\ref{sec:results} presents the derived correction factors. Section~\ref{sec:Discussion} discusses the results. Section~\ref{sec:Conclusions} summarises the main conclusions.

\FloatBarrier

\section{Data Observations}
\label{sec:DataObservations}
In this work, we used observations from the three hyperspectral instruments, which provide high-spatial-resolution Mars observations at these wavelengths. The primary dataset consists of three HS-H images of Mars acquired on 12 March, 2025, during the Hera spacecraft flyby, from which  47 spectra were extracted across 21 surface regions (see Fig.~\ref{fig:appendix2}). These were compared against 47 spectra from OMEGA/Mars Express and  47 spectra from CRISM/MRO, selected to match the same surface regions. The instruments and their respective datasets are described in detail in the following subsections.

\subsection{HyperScout-H}
\label{subsec:HSH}
The HS-H instrument is a miniaturized snapshot hyperspectral imager developed by Cosine Remote Sensing BV, built on the HyperScout 2 platform originally designed for Earth observation \citep{EspositoMarchi2015, EspositoMarchi2019}. The instrument has a mass of 5.145 kg and a volume of 238 $\times$ 190 $\times$ 277~mm$^3$. The telescope is based on a Three-Mirror Anastigmat configuration with an aperture diameter of 10.3125 mm and a focal length of 41.25 mm, providing a wide field of view of approximately 15.5$^\circ$ $\times$ 8.3$^\circ$ in paraxial approximation (15.9$^\circ$ $\times$ 9.9$^\circ$ accounting for optical distortions, \citet{Popescu2025}). \\

The detector is an AMS CMV2000 Silicon Complementary Metal-Oxide-Semiconductor (CMOS) array coated with pixelated Fabry-Pérot interference filters, with a pixel pitch of 5.5 $\mu$m and a total format of 2048 $\times$ 1088 subpixels. The spectral separation is achieved through a macropixel architecture, in which each macropixel consists of a 5 $\times$ 5 matrix of subpixels, each covered by a different narrow-band filter. This yields a total of 25 spectral bands spanning the 0.65–0.95 $\mu$m wavelength range, with median central wavelengths ranging from 0.661 to 0.952 $\mu$m and band FWHM values between 0.008 and 0.022 $\mu$m. The effective detector format in macropixels is 409 $\times$ 217. The analog-to-digital converter (ADC) has a 12-bit resolution. At a distance of 1~km, the  spatial sampling is 13.2~cm per subpixel and 66~cm per macropixel \citep{Popescu2025}. The filter profiles exhibit out-of-band ripples, with an average in-band to out-of-band ratio of 0.7, which must be accounted for when comparing HS-H data with spectra from other instruments. \\

The spectral response of each channel is described by its optical transfer function, which includes the transmission profile of the Fabry-Pérot interference filter and the quantum efficiency of the CMOS detector (Eq. \ref{eq:transfer}). For a given spectrum $S(\lambda)$, the signal recorded in channel i is computed as:

\begin{equation}
S^i_{\rm HS-H} = \frac{\int F^i(\lambda) S(\lambda) \lambda \, d\lambda}{\int F^i(\lambda) \lambda d\lambda}
\label{eq:transfer}
\end{equation}
where F($\lambda$) is the transfer function of channel i. The sensitivity factors for each channel, derived from pre-flight radiometric calibration performed at the laboratories of the European Space Research and Technology Centre (ESTEC), are reported in \citet{Popescu2025}. The in-flight sensitivity factors, which are the ones actually applied to the data analyzed here, are those of \citet{Prodan2026}.

\subsection{Pre-flight and In-flight Calibration}
\label{subsec:calibration}
The pre-flight calibration of HS-H was performed in two independent campaigns, by cosine Remote Sensing BV and at ESA/ESTEC, with results found to be consistent within $\pm$2\% \citep{Popescu2025}. The calibration procedure included dark-current characterization, flat-field correction, and bad-pixel mapping — identifying only 12 individual hot pixels and 3 malfunctioning detector rows, which were replaced with NaN values in the calibrated images.  Radiometric calibration was performed using a 0.5~m integrating sphere, deriving the sensitivity factor for each channel. The relative uncertainty of these factors is 0.06\% per channel, computed as the dispersion among the pixels of the same channel; it does not include the 0.25\% peak-to-peak stability of the lamp output, and it quantifies the repeatability of the laboratory measurement rather than the accuracy of the absolute radiometric scale \citep{Popescu2025}. Geometric distortion was quantified using a checkerboard target and modeled by two-dimensional third-order polynomial correction functions $f_x(x_d, y_d)$ and $f_y(x_d, y_d)$, such that:
\begin{equation}
f_x(x_d, y_d) = \Delta x, \quad f_y(x_d, y_d) = \Delta y
\label{eq:distortion}
\end{equation}
where $\Delta x = x_u - x_d$ and $\Delta y = y_u - y_d$ represent the pixel displacement corrections along each 
axis, and $(x_d, y_d)$, $(x_u, y_u)$ are the distorted and undistorted coordinates respectively \citep{Prodan2026}. 
When this pre-flight model is applied to the in-flight star field observations, it leaves RMS residuals of 14.9 and 12.3~pixels along the X and Y axes \citep{Prodan2026}.  \\  

The in-flight calibration campaign \citep{Prodan2026} was designed to refine the detector characterization under actual operational conditions, accounting for launch stresses, thermal cycling, radiation exposure, and zero-gravity effects that cannot be fully replicated on the ground. The bias pattern was found to be stable at $\sim$151~DN, with a standard deviation of 3.4~DN and a readout noise of $\sim$8~DN. Dark current at exposure times $\leq$100~ms was negligible, confirming that no dark correction is required for the short integrations planned during the asteroid phase. Detector linearity was confirmed with $R^2 = 0.9997$. The PSF was characterized from observations of a bright star (Vega), yielding FWHM values of 1.54 and 1.60 pixels along the two detector axes, consistent with the pre-flight value of 1.6 pixels. Geometric distortion was corrected using 71 star field frames with 312 star detections, achieving RMS residuals of 0.36 and 1.87 pixels along the X and Y axes — a significant improvement over the pre-flight correction \citep{Prodan2026}. \\

Radiometric calibration was performed using  stellar observations, from which 367 stellar profiles were extracted. The resulting in-flight sensitivity constants differ by approximately 10\% from the pre-flight values, attributed to the lower operating temperature in space ($- 12^{\circ}$C versus $+20^{\circ}$C in the laboratory). A preliminary comparison with CRISM data over five Martian surface regions revealed systematic channel-dependent offsets, motivating the extended cross-calibration presented in this work. The calibrated images produced by this pipeline constitute the starting point of our analysis \citep{Prodan2026}. \\

The uncertainty of the HS-H spectra combines three contributions. The first is the detector noise. Since each of the 25 spectral samples of a macropixel is provided by a single sub-pixel in a single exposure, the band-to-band scatter of an individual macropixel spectrum is 3--5\%, measured on homogeneous areas of the Martian images and scaled with the signal level of each region, corresponding to a signal-to-noise ratio between 20 and 33 per spectral sample. This term decreases as $N^{-1/2}$ when $N$ macropixels are averaged, and falls to 0.3--0.5\% for the values of $N$ between 90 and 105 that cover the regions considered in this work. The second contribution is the relative spectral uncertainty of the in-flight radiometric constants, which ranges from 0.3\% to 0.9\% depending on the channel and does not decrease with averaging; referred to the normalization channel at 0.806~\textmu m it is 0.74\% on average. The third is the absolute radiometric uncertainty of 1.3\%, which is common to all 25 channels: it acts as a scale factor on the whole spectrum, it does not contribute to the band-to-band scatter, and it cancels exactly in the normalized spectra and in the normalized correction factors. Combining the first two terms, the uncertainty of a region-averaged HS-H spectrum is 0.9--1.0\% per band, and this is the value adopted for the HS-H spectra in the following. This estimate agrees with the data. We fitted a smooth polynomial continuum to each region-averaged spectrum and measured the scatter of the 25 points around it. The scatter is 0.6--1.3\% for HS-H, against 0.3--0.4\% for OMEGA and CRISM over the same regions. It is an upper limit, because there can be contributions from actual spectral variations within the FOV. The flat-field correction is already applied to the calibrated products, and any residual is contained in the band-to-band scatter given above.

\subsection{HS-H Dataset}
\label{subsec:HSH_dataset}
\begin{figure*}[!t]
\includegraphics[width=\textwidth]{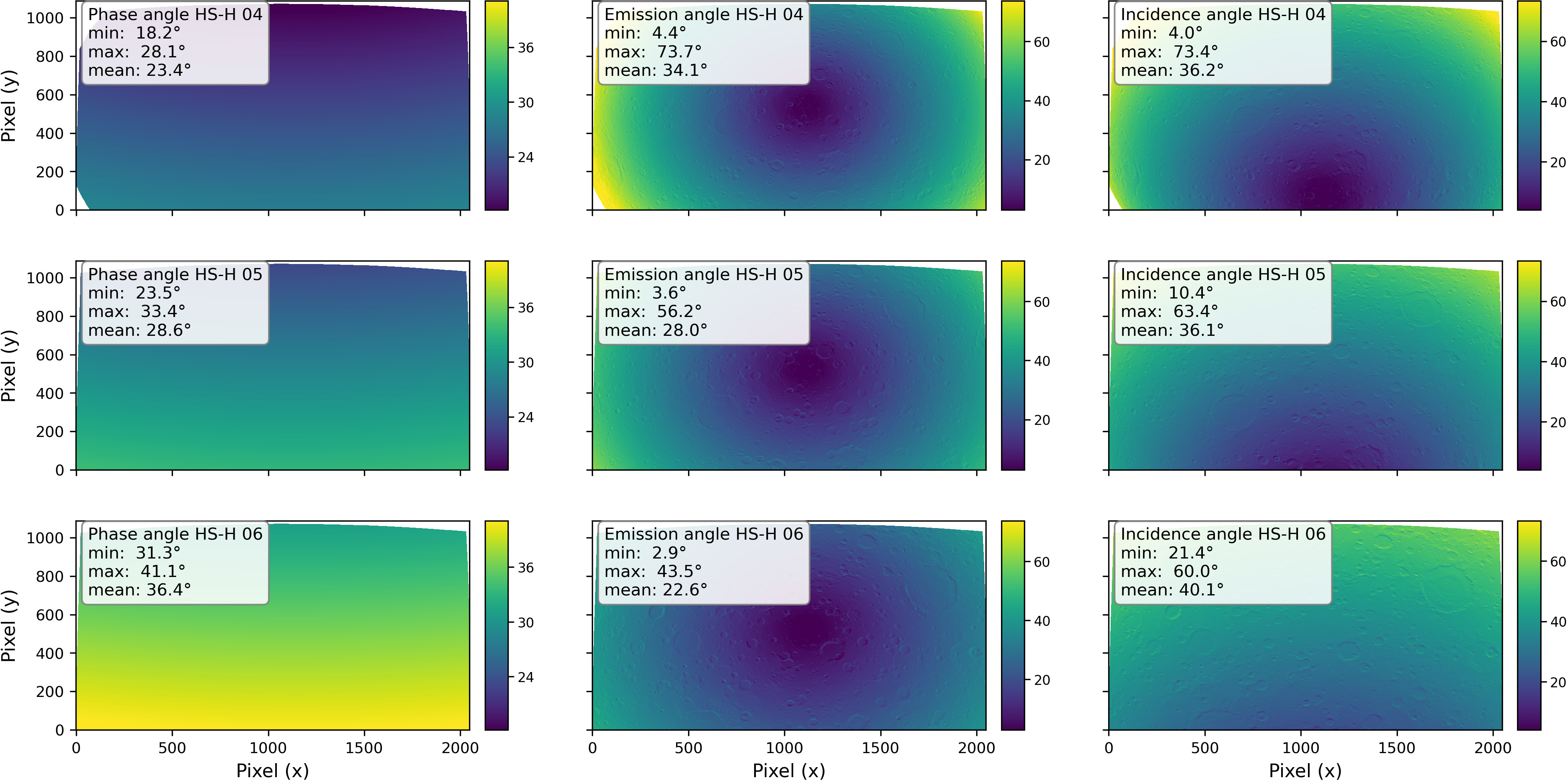}
\caption{Spatial distribution of the phase angle (left), emission angle (center), and incidence angle (right) for the three HS-H images used in this work (rows from top to bottom: HS-L0-FE-025D-04-01, HS-L0-FE-025D-05-01, HS-L0-FE-025D-06-01). Values are given in degrees. Invalid border pixels are shown in white. These maps illustrate the illumination conditions for the observations analyzed in this work.}
\label{fig:HH_geometry}
\end{figure*}

In this work, we used three HS-H images of the Martian surface, acquired on 12 March 2025, between 12:15:45.944 and 12:29:45.979 UTC, with an exposure time of 0.01~s (Table~\ref{tab:HH_obs}). The images are identified as HS-L0-FE-025D-04-01, HS-L0-FE-025D-05-01, and HS-L0-FE-025D-06-01. During this acquisition window, the distance from Mars ranged from 20\,924.8~km to 14\,521.4~km, corresponding to  a spatial sampling between 13.8 and 9.6~km/ macropixel. These three images were selected over the earlier ones acquired during the flyby because of their  finer spatial sampling, which is better suited for comparison with OMEGA and CRISM data. The spatial distribution of the phase, emission, and incidence angles across the three images is shown in Fig.~\ref{fig:HH_geometry}. \\

The images were processed using the existing calibration pipeline described in Section~\ref{subsec:calibration}, including bias subtraction, flat-field correction, bad-pixel interpolation, and geometric distortion correction. Radiometric calibration was performed by  dividing the  digital number counts (DN)  by the in-flight sensitivity factors derived by \citet{Prodan2026} (Table~5 therein), thereby converting the data to units of  $W~m^{-2}~sr^{-1}~nm^{-1}$. The resulting radiance images were then divided by the band-averaged solar irradiance values \citep{Prodan2026} to obtain reflectance factor (I/F) images. For each HS-H image, co-registered maps of latitude, longitude, phase angle, emission angle, and incidence angle were available as separate FITS files produced by the in-flight calibration pipeline \citep{Prodan2026}. These maps share the same pixel grid as the HS-H science images, allowing the geometric conditions to be retrieved pixel-by-pixel. The per-pixel geometric information was used to characterize the observing conditions of each selected surface region, as described in Section~\ref{subsubsec:HH_extraction}.

\begin{table*}[!t]
\centering
\caption{Observational parameters of the three HS-H images used in this work, acquired on 12 March 2025 during the Hera Mars flyby. Latitude and longitude are given as min/max values over the valid pixels of each image. Phase, emission, and incidence angles are given as mean values over the valid pixels of each image. The spatial sampling is computed from the spacecraft--Mars distance and the macropixel instantaneous field of view of 0.66~mrad.}
\label{tab:HH_obs}
\resizebox{\textwidth}{!}{%
\begin{tabular}{lccccccccc}
\hline\hline
Image ID & UTC Start & $t_\mathrm{exp}$ & Distance &  Spatial sampling & Lat. range & Lon. range & Phase & Emission & Incidence \\
         &           & (s)              & (km)     & (km~Mpx$^{-1}$) & (°) & (°) & (°) & (°) & (°) \\
\hline
HS-L0-FE-025D-04-01 & 12:15:45.9 & 0.01 & 20924.8 &  13.8 & $-38.24$ / $29.84$ & $-28.98$ / $81.08$ & 23.40 & 34.15 & 36.23 \\
HS-L0-FE-025D-05-01 & 12:22:46.9 & 0.01 & 17610.2 &  11.6  & $-36.89$ / $18.84$ & $-9.72$  / $70.94$ & 28.62 & 27.95 & 36.08 \\
HS-L0-FE-025D-06-01 & 12:29:45.9 & 0.01 & 14521.4 &  9.6   & $-37.06$ / $6.69$  & $3.67$   / $64.55$ & 36.36 & 22.58 & 40.06 \\
\hline
\end{tabular}%
}
\end{table*}

\FloatBarrier

\section{OMEGA and CRISM}
\label{sec:OMEGA_CRISM}

\subsection{OMEGA and CRISM data sets}
\label{subsec:instruments}

\bigskip

The CRISM data products used in this work are the VNIR Hyperspectral Reduced Data Records (VRDR),  which are map-projected mosaics assembled from several strips of VNIR hyperspectral mapping observations and distributed on a global grid of tiles of approximately $5^\circ \times 5^\circ$ \citep{Viviano2014}. Their processing includes a Lambertian photometric correction and a volcano-scan atmospheric correction. Empirical corrections are also applied to remove spectral smile. The VRDR products cover the visible/near-infrared range (0.37--1.02~$\mu$m) and provide spectra already expressed as reflectance factor (I/F divided by the cosine of the incidence angle), at the full spectral resolution of the VNIR detector, which is required in order to convolve them with the HS-H filter transfer functions. Because each tile is a mosaic of strips acquired at different epochs, the selection in solar longitude described in Sect.~\ref{subsec:dataretrival} was applied at the pixel level rather than to the product as a whole.

\begin{figure}[!ht]
    \centering
    \includegraphics[width=\linewidth,height=0.80\textheight,keepaspectratio]{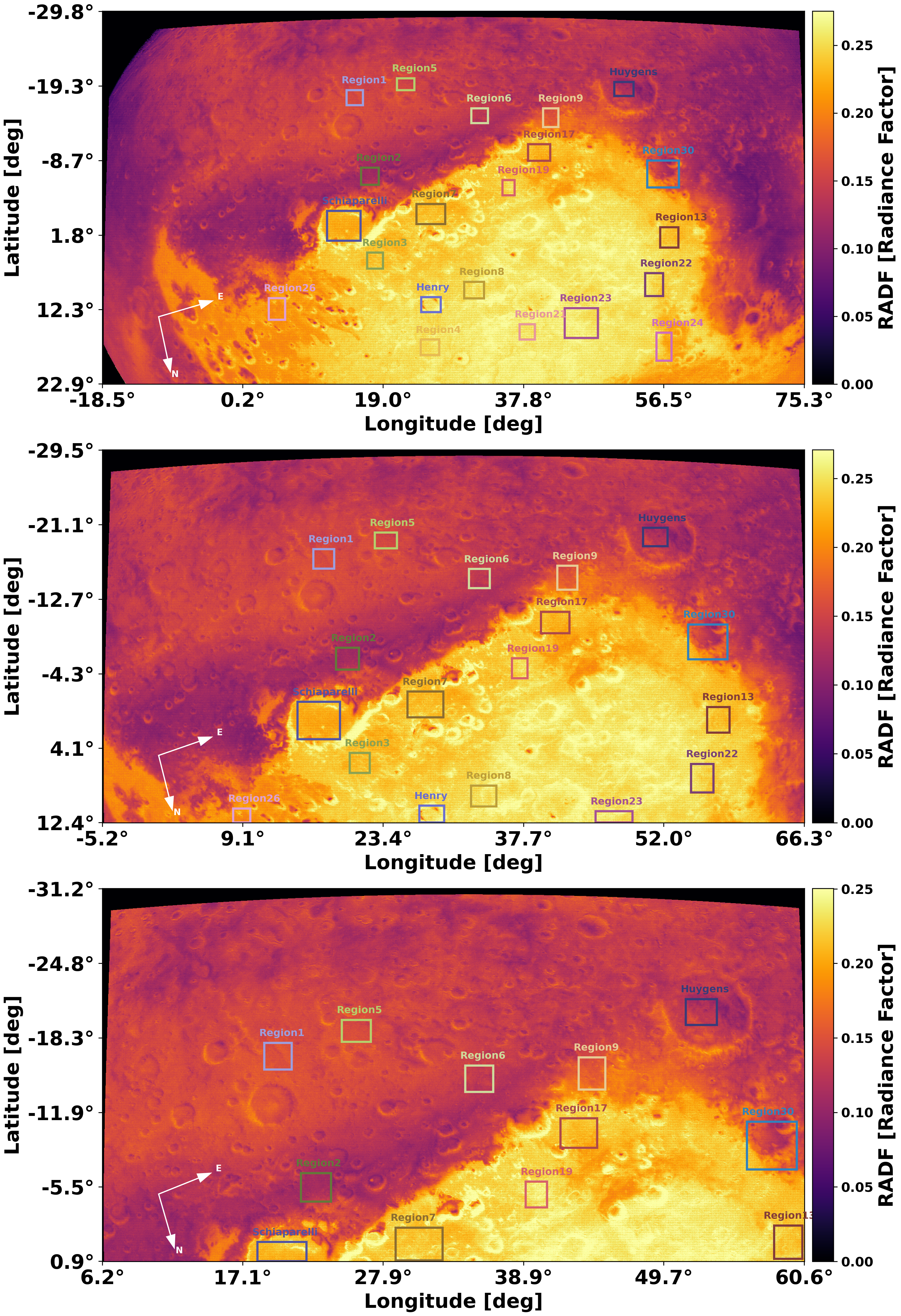}
    \caption{The three HS-H images used in this work (HS-L0-FE-025D-04-01, HS-L0-FE-025D-05-01, and HS-L0-FE-025D-06-01, from top to bottom), with the 21 selected surface regions overlaid as colored rectangles. Latitude and longitude are shown on the axes. The color of each rectangle identifies the region type: craters (Huygens, Schiaparelli, Henry) and additional surface regions (Region~1--Region~30). The region names are those used in the original selection and are therefore not consecutive, since some regions were discarded by the selection criteria.}
    \label{fig:HH_regions}
\end{figure}

The OMEGA and CRISM reference datasets were selected to match the surface area covered by the HS-H image HS-L0-FE-025D-04-01, which provides the widest geographic coverage among the three images used in this work (Table~\ref{tab:HH_obs}).
Within the image's latitude and longitude ranges, 21 surface regions were identified for spectral extraction. The region names used throughout this work are those assigned during the initial selection and are therefore not consecutive, since some of the regions were subsequently discarded by the selection criteria described below. The 21 selected regions are shown in Fig.~\ref{fig:HH_regions} overlaid on the three HS-H images.

\begin{figure*}[!t]
\centering


\begin{subfigure}{0.32\textwidth}
    \centering
    \includegraphics[width=\linewidth]{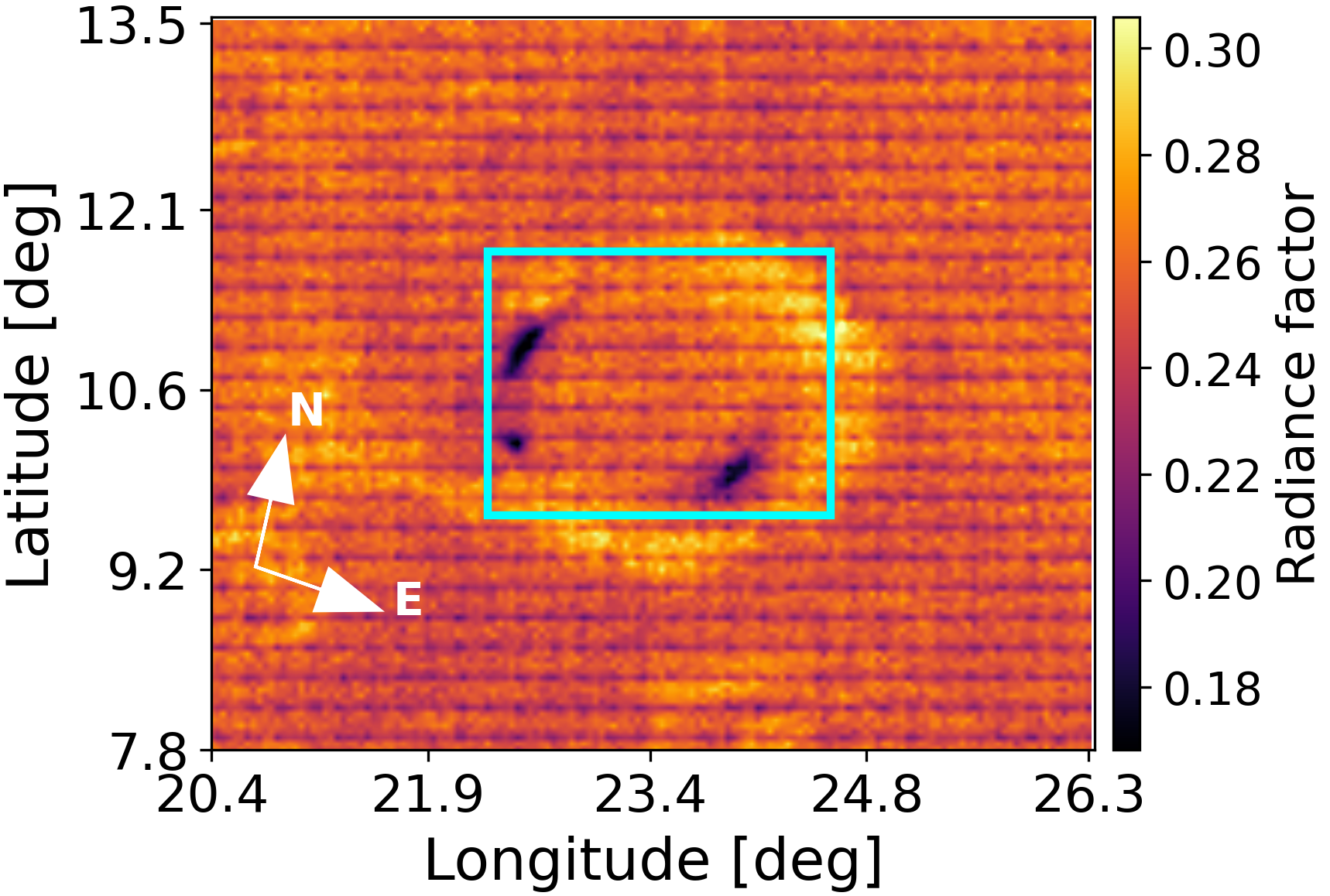}
\end{subfigure}
\hfill
\begin{subfigure}{0.32\textwidth}
    \centering
    \includegraphics[width=\linewidth]{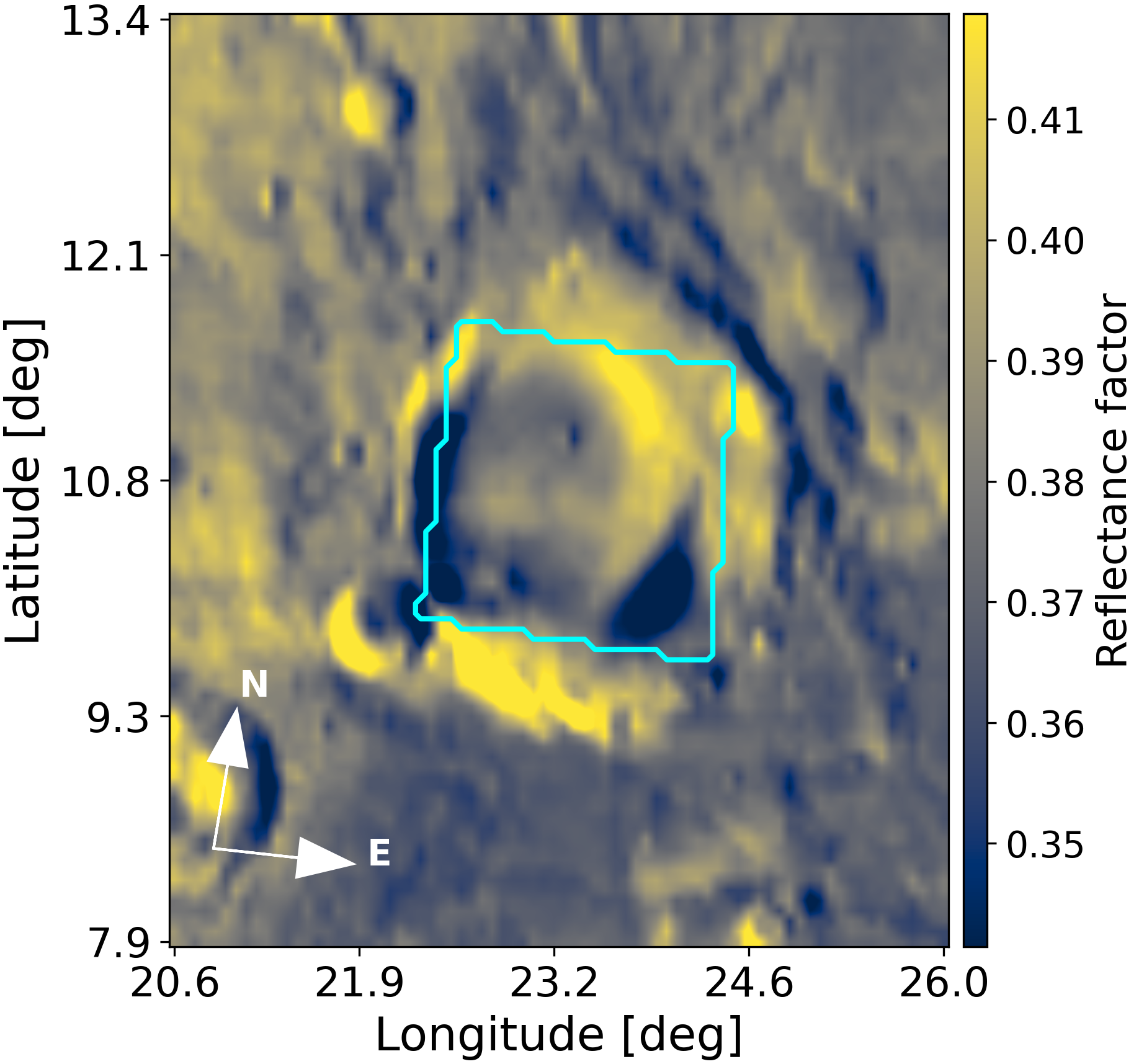}
\end{subfigure}
\hfill
\begin{subfigure}{0.32\textwidth}
    \centering
    \includegraphics[width=\linewidth]{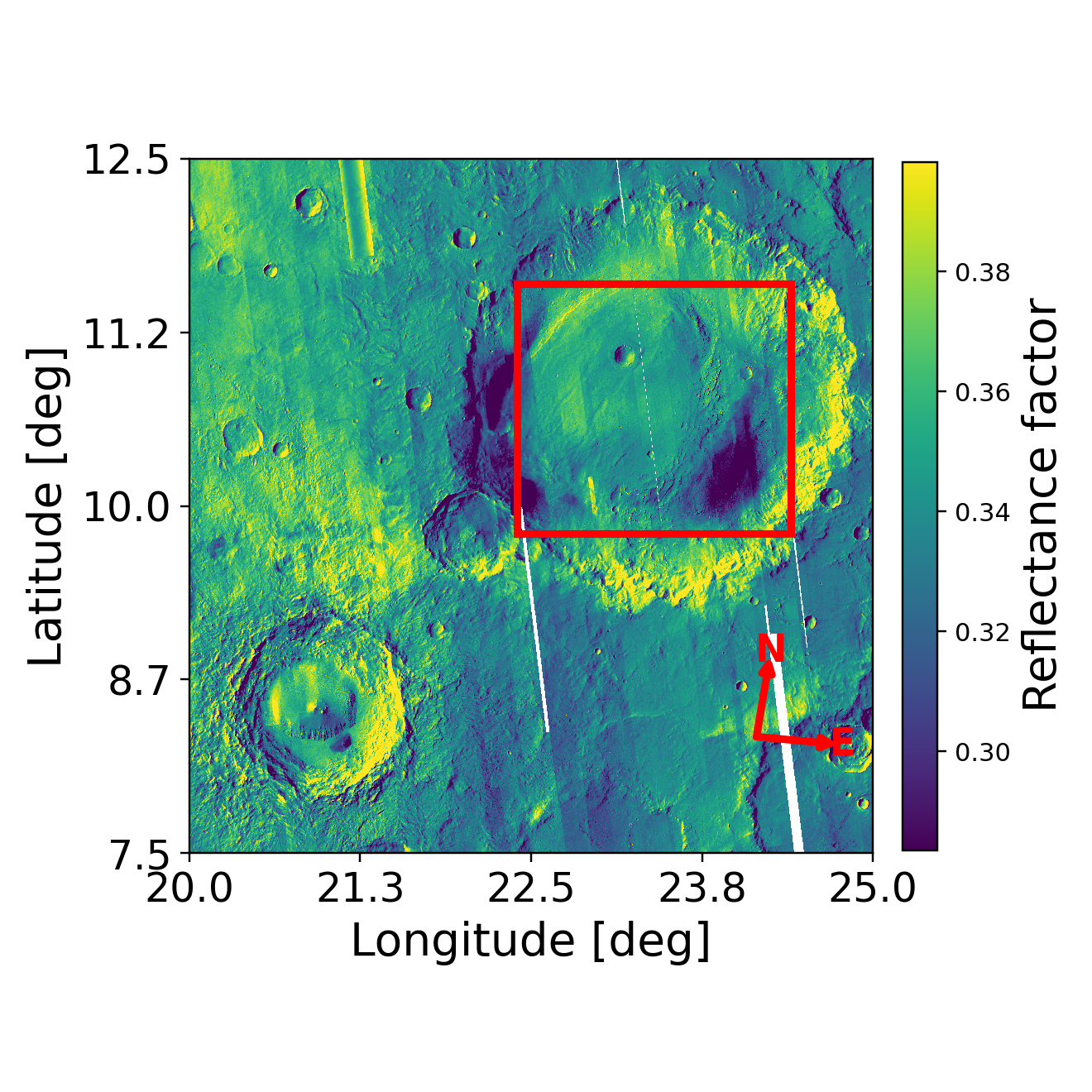}
\end{subfigure}

\vspace{0.5cm}


\begin{subfigure}{0.32\textwidth}
    \centering
    \includegraphics[width=\linewidth]{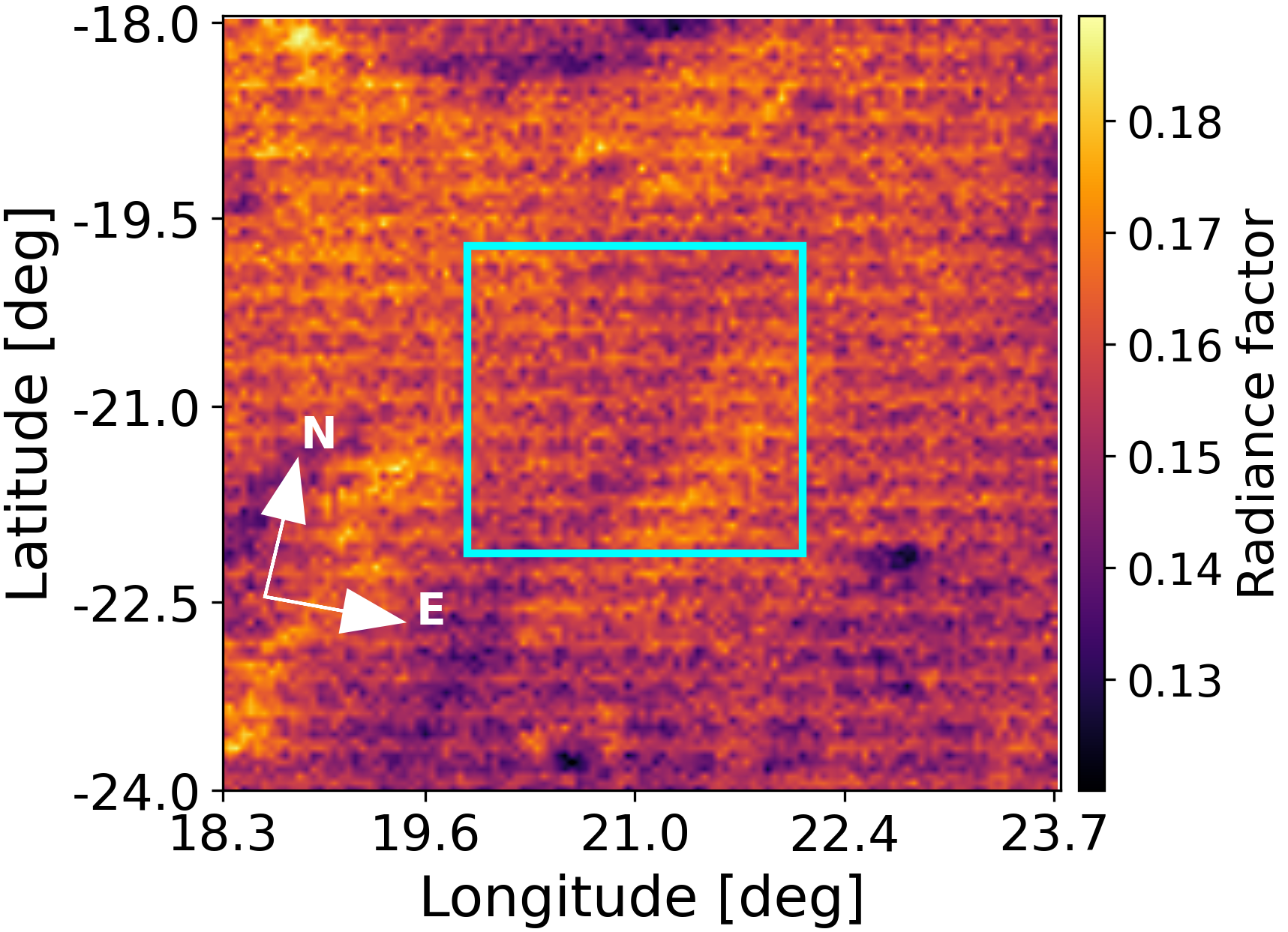}
\end{subfigure}
\hfill
\begin{subfigure}{0.32\textwidth}
    \centering
    \includegraphics[width=\linewidth]{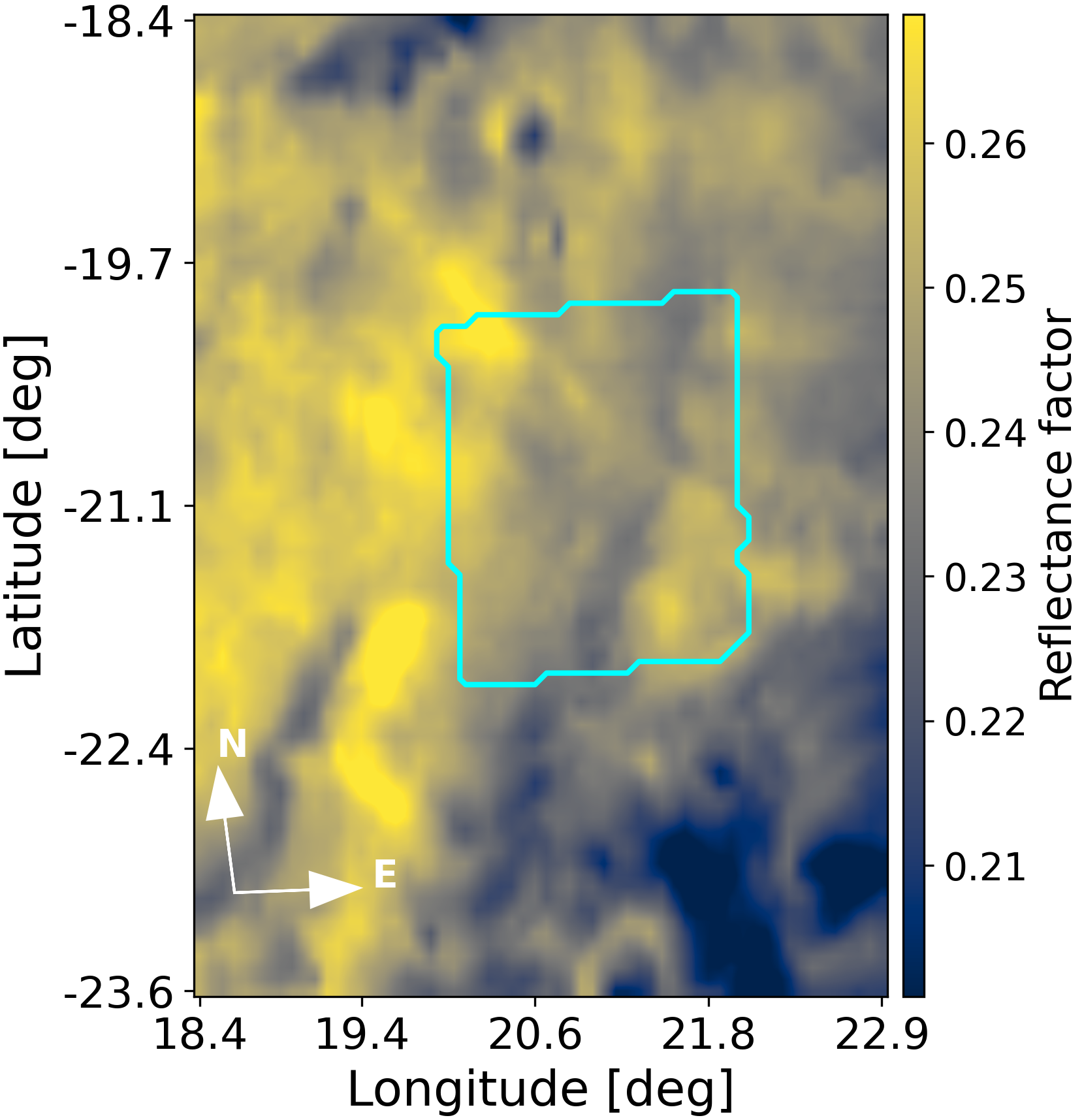}
\end{subfigure}
\hfill
\begin{subfigure}{0.32\textwidth}
    \centering
    \includegraphics[width=\linewidth]{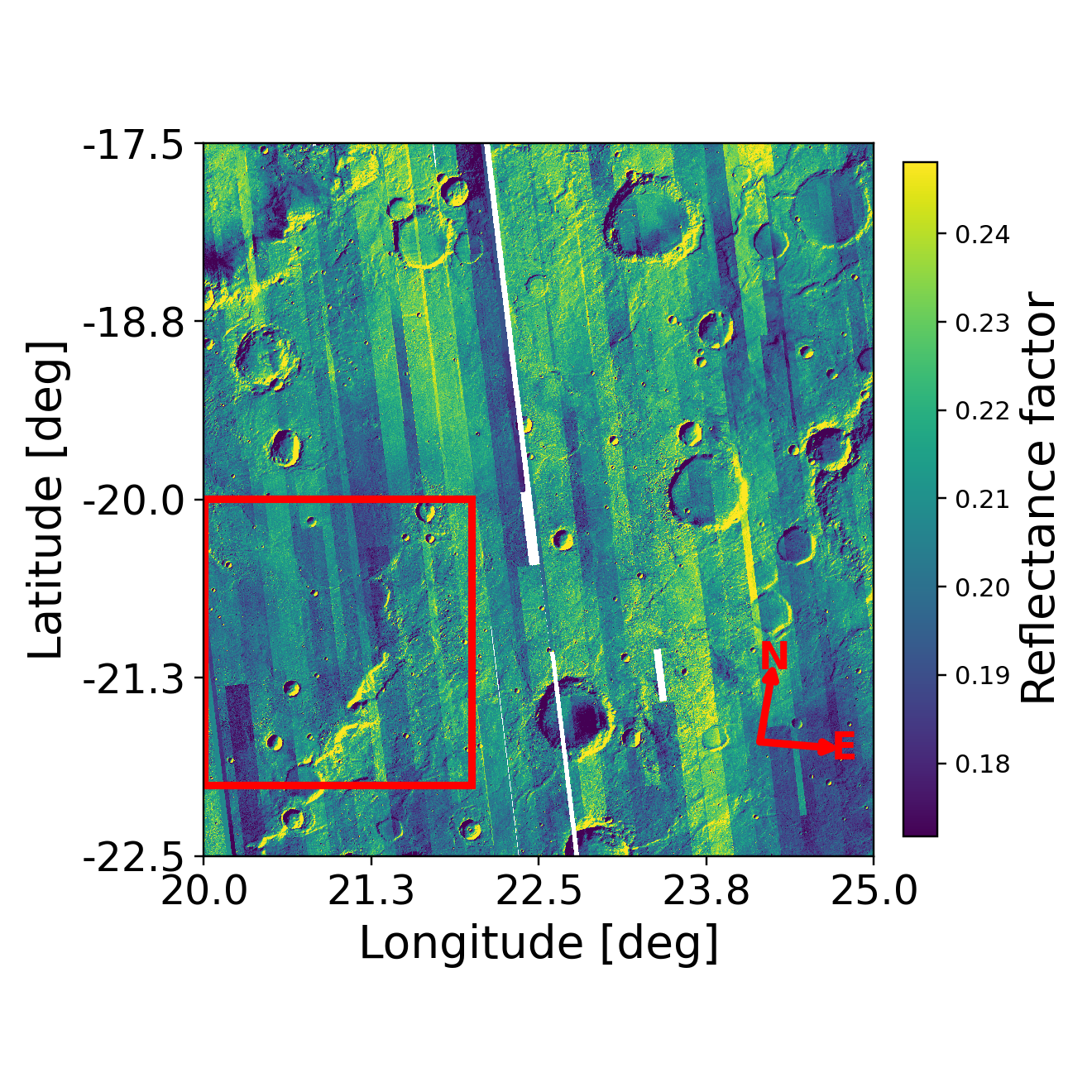}
\end{subfigure}

\vspace{0.5cm}


\begin{subfigure}{0.32\textwidth}
    \centering
    \includegraphics[width=\linewidth]{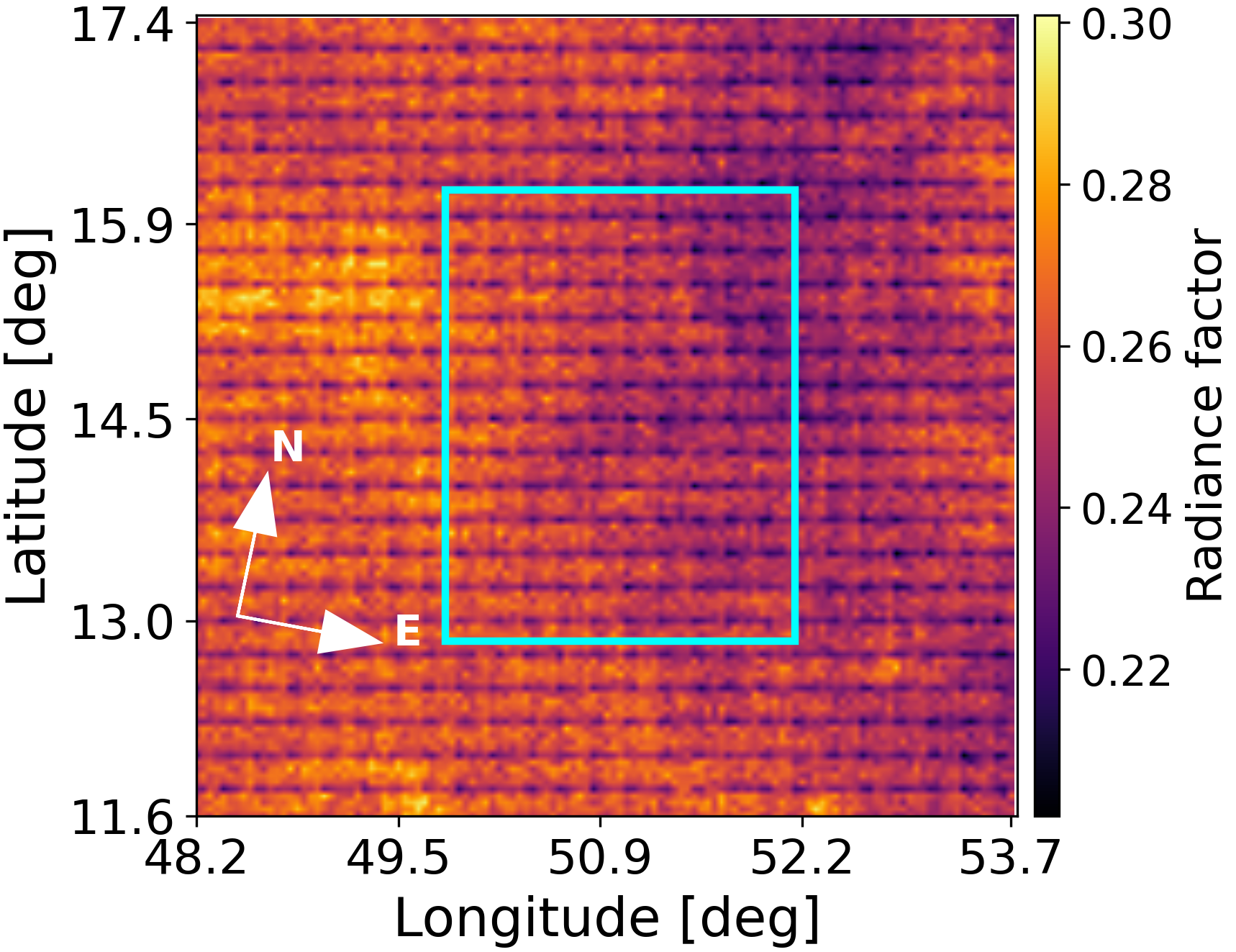}
\end{subfigure}
\hfill
\begin{subfigure}{0.32\textwidth}
    \centering
    \includegraphics[width=\linewidth]{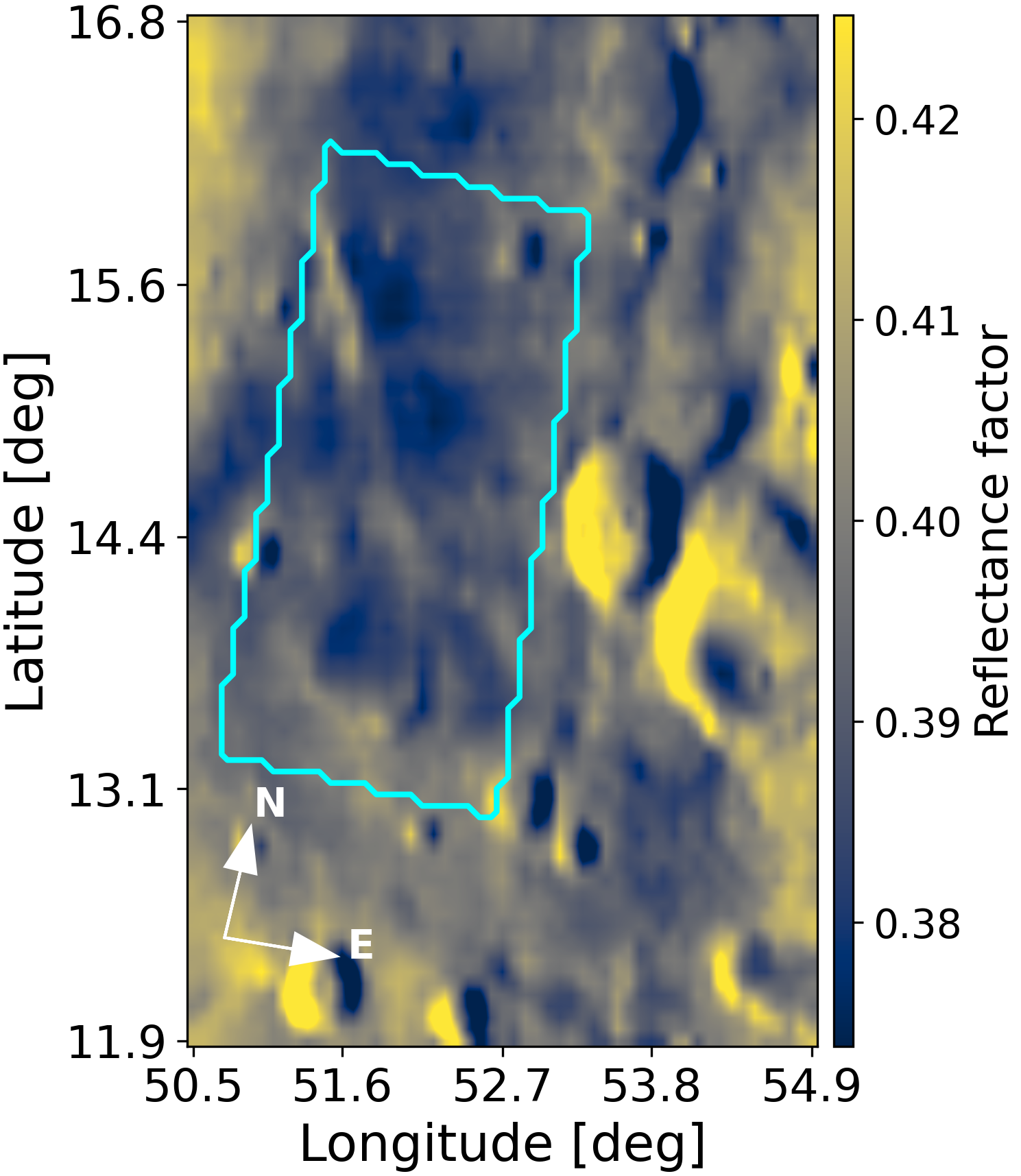}
\end{subfigure}
\hfill
\begin{subfigure}{0.32\textwidth}
    \centering
    \includegraphics[width=\linewidth]{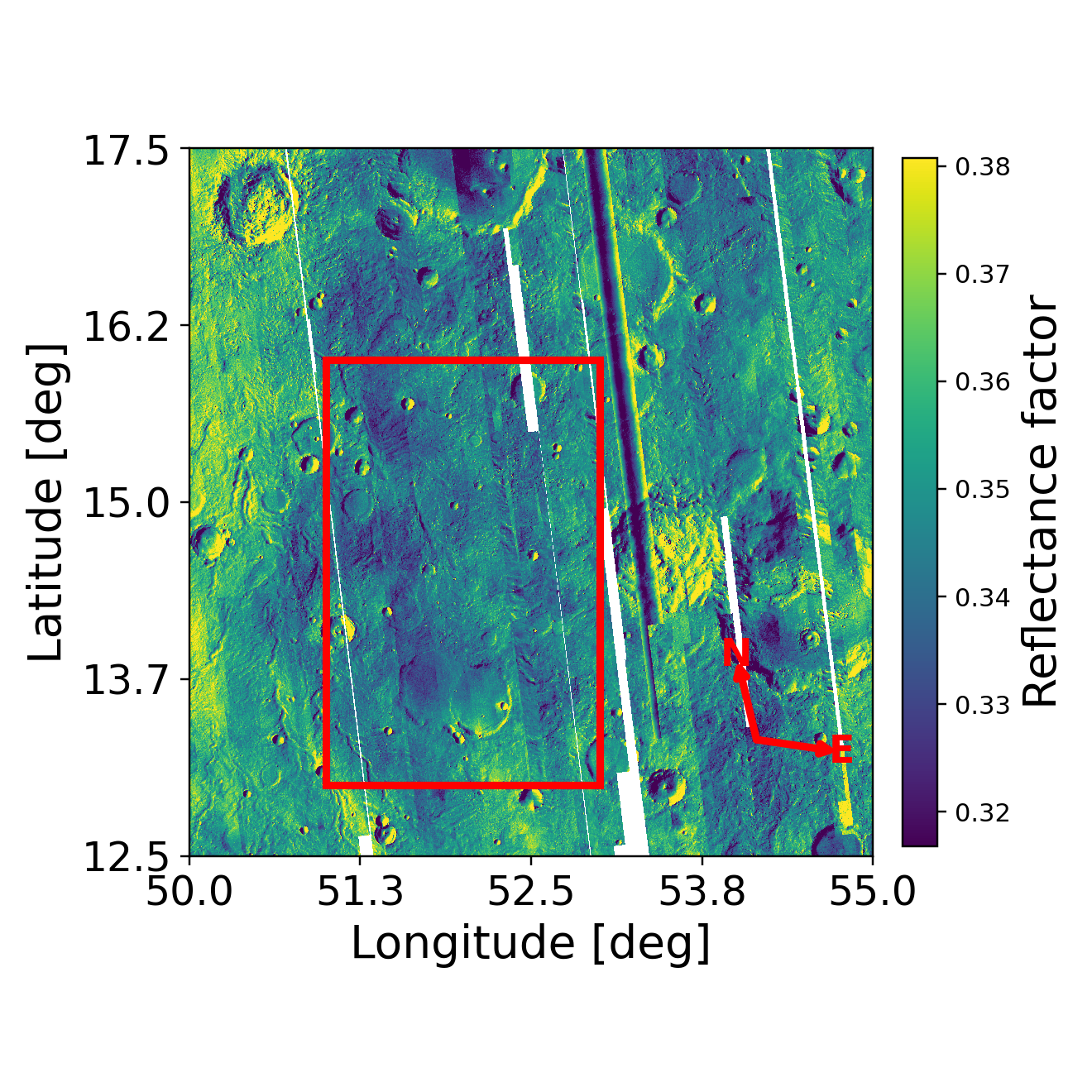}
\end{subfigure}

\caption{Representative examples of three selected surface regions as observed by HS-H (left column), OMEGA (center column), and CRISM (right column). From top to bottom: Henry crater, Region~1 ( dark terrain), and Region~22 ( bright terrain). The cyan contour (HS-H and OMEGA) and the red rectangle (CRISM) mark the region used for spectral extraction. North is up and East is to the right in all panels. The HS-H panels show the radiance factor $I/F$, the OMEGA and CRISM panels the reflectance factor as provided by the data products; all spectra used in the analysis are reflectance factors, that is $I/F$ divided by the cosine of the solar incidence angle (Sect.~\ref{subsubsec:HH_extraction}). The OMEGA images are ORB0558\_4 ($L_s = 53.1^\circ$), ORB1505\_5 ($L_s = 178.5^\circ$) and ORB0554\_4 ($L_s = 52.6^\circ$), all in northern spring or summer. The CRISM mosaics are shown for context only; the spectra were extracted from the pixels satisfying the $L_s$ criterion.}

\label{fig:region_comparison}

\end{figure*}

The regions were selected visually on the HS-H image after applying a flux color map to distinguish bright and dark terrains. Representative examples of three selected regions — Henry crater, Region~1 ( dark terrain), and Region~22 ( bright terrain) — as observed by HS-H, OMEGA, and CRISM are shown in Fig.~\ref{fig:region_comparison}. The selection was designed to sample a variety of surface types, crater floors and rims (Huygens, Schiaparelli, and Henry craters), bright and dark terrains, image border regions, and additional areas distributed across the scene, ensuring a representative coverage of the observed surface. OMEGA observations covering these 21 regions were retrieved from the ESA Planetary Science Archive\footnote{\url{https://psa.esa.int/psa/}} in the form of spectral image cubes (.QUB) and their associated navigation files (.NAV). A total of  37 OMEGA observations from  26 unique orbits were selected, with multiple orbits per region to increase the statistical robustness of the comparison. Since a single observation can cover more than one region, these observations yield a larger number of region--observation pairs. CRISM observations covering the same 21 regions were retrieved from the PDS Geosciences Node\footnote{\url{https://pds-geosciences.wustl.edu/missions/mro/crism.htm}},  in the form of VNIR products. The spectral extraction procedure and the criteria used to select the final set of  47 spectra from both instruments are described in Section~\ref{subsec:dataretrival}. The OMEGA observations used in this work have a spatial sampling between 0.8 and 4.0~km per pixel, with a median of 1.9~km, while the CRISM VRDR tiles are map-projected at 256 pixels per degree, corresponding to about 230~m per pixel.

The section is organized as follows. Sections~\ref{subsubsec:HH_extraction}, \ref{subsubsec:OMEGA_extraction}, and \ref{subsubsec:CRISM_extraction} describe the spectral extraction procedures for the three instruments. Section~\ref{subsec:preprocessing} describes the spectral preprocessing, including the application of the HS-H filter transfer functions. Section~\ref{subsec:dataretrival} presents the criteria adopted to select the OMEGA and CRISM observations entering the comparison.
Sections~\ref{subsec:CF_norm} and~\ref{subsec:CF_flux} present the correction factors derived from the normalized and absolute flux spectra, respectively. Section~\ref{subsec:repeatability} assesses the 
repeatability and reliability of the results across the three HS-H images. Finally, Section~\ref{subsec:geometry} analyzes the dependence of the absolute flux correction factor on the observing geometry parameters and on the surface albedo.

\subsection{Data retrieval}

\subsubsection{HS-H Spectral Extraction}
\label{subsubsec:HH_extraction}

Each HS-H image was first calibrated in wavelength using the per-pixel central-wavelength calibration map \citep{Popescu2025}, which assigns an effective central wavelength to each subpixel. The macropixel architecture of HS-H, in which each 5~$\times$~5 matrix of subpixels samples 25 distinct spectral bands, is described in detail in \citet{Popescu2025}. Each image was also registered with co-registered maps of latitude, longitude, phase angle, emission angle, and incidence angle, produced by the in-flight calibration pipeline \citep{Prodan2026}, allowing the geometric conditions to be retrieved at the subpixel level. \\

For each of the 21 surface regions defined in Section~\ref{sec:OMEGA_CRISM}, a geographic mask was constructed by selecting all subpixels whose latitude and longitude fell within the boundaries of the region. The selected subpixels were then extended to include complete macropixels, any macropixel containing at least one selected subpixel was fully included in the extraction, ensuring spectral completeness across all 25 bands. When this procedure produced more than one disconnected area, only the largest connected component was retained. \\

Prior to spectral extraction, the flux of each subpixel was corrected for illumination geometry by dividing by the cosine of the local incidence angle, retrieved from the per-pixel incidence angle map. 
This correction places the HS-H data on the same photometric convention as OMEGA and CRISM, both of which provide data already expressed as reflectance factor, that is $I/F$  divided by the cosine of the incidence angle \citep{Mustard2007, Viviano2014}. For each spectral band, the mean and median fluxes were computed over all valid subpixels within the region, yielding the mean and median spectra of the surface region as seen by HS-H. The number of subpixels contributing to each region varies with region size, which was defined to match the spatial extent of the corresponding OMEGA observation. Regions that fell outside the spatial coverage of a given HS-H image were excluded from the analysis for that image. The resulting spectra were saved as two-column text files containing the effective wavelength and flux value for each of the 25 bands. The uncertainty of each band of the resulting region spectrum was propagated following the error budget of Sect.~\ref{subsec:calibration}, combining the detector noise, which decreases with the number of macropixels averaged, with the relative spectral calibration uncertainty, which does not. \\

\subsubsection{OMEGA Spectral Extraction}
\label{subsubsec:OMEGA_extraction}

OMEGA observations were processed using the \textsc{omegapy} Python library \citep{Stcherbinine2024}, a re-implementation of the official OMEGA IDL pipeline. For each observation, the level~1B binary files (.QUB and .NAV) downloaded from the ESA Planetary Science Archive were loaded and processed through the thermal and atmospheric correction function (\texttt{corr\_therm\_atm}), which applies the volcano-scan atmospheric correction \citep{Stcherbinine2024} and removes the thermal emission contribution at longer wavelengths. The output is the reflectance factor cube (\texttt{cube\_rf}), defined as the radiance factor divided by the cosine of the incidence angle, consistent with the photometric convention described in \citet{Mustard2007} and with the correction applied to the HS-H data described in Section~\ref{subsubsec:HH_extraction}. \\

For each of the 21 surface regions, all available OMEGA observations covering the region were retrieved and processed. Each OMEGA observation is identified by  an identifier of the form ORBXXXX\_X, where ORBXXXX denotes the Mars Express orbit, that is a single passage over the Martian surface, and the suffix identifies the individual data cube acquired during that orbit. A single orbit can therefore provide more than one observation. For each region, a latitude-longitude bounding box was defined, and all ORBs whose spatial coverage overlapped with that box were selected. The median reflectance spectrum was extracted from all valid pixels falling within the bounding box for each ORB. In cases where an ORB covered only part of the selected region, only the pixels within the available overlap were used. Since a single ORB can overlap with multiple regions and a single region can be covered by multiple ORBs, the total number of ORB--region pairs (47, from  26 unique orbits) exceeds the number of regions. 

The same latitude--longitude bounding boxes were used to extract the corresponding CRISM spectra, as described in Section~\ref{subsubsec:CRISM_extraction}. For each OMEGA observation, a geographic mask was constructed by selecting all pixels whose latitude and longitude fell within the boundaries of the corresponding surface region, as defined in the mapping file. The mean and median reflectance spectra were then computed over all selected pixels, yielding two continuous spectra covering the full OMEGA wavelength range. Observations with no valid pixels in the selected region or that were affected by saturation or other data quality issues were excluded. The resulting spectra were saved as two-column text files containing the OMEGA wavelength array and the corresponding mean or median reflectance value. Since the OMEGA  spectral samples beyond 0.930~$\mu$m are affected by contamination from high grating orders and are therefore not delivered in the data products, leaving a gap between approximately 0.930 and 1.040~$\mu$m in the retrieved spectra, the last three HS-H bands (0.934, 0.944, and 0.952~$\mu$m) were consequently excluded from the OMEGA comparison in the subsequent analysis. A summary of all OMEGA observations used, with their associated surface regions, coordinate boundaries, and observing geometry, is provided in Table~\ref{tab:geometry_appendix} in the Appendix.

\subsubsection{CRISM Spectral Extraction}
\label{subsubsec:CRISM_extraction}

CRISM observations covering the 21 selected surface regions were processed using a dedicated Python pipeline. The input data consisted of VNIR Hyperspectral Reduced Data Records (VRDRs) retrieved from the PDS Geosciences Node, covering the 0.37--1.02~$\mu$m wavelength range. The VRDR products include the  corrections described in Section~\ref{subsec:instruments}, including Lambertian photometric correction and volcano-scan atmospheric correction \citep{McGuire2009} (a standard empirical correction for the Martian atmosphere based on observations of Olympus Mons at multiple altitudes), and provide spectra already expressed as reflectance factor (I/F divided by the cosine of the incidence angle), consistent with the photometric convention used for OMEGA and HS-H. No additional corrections were applied beyond those already included in the data products. \\

For each observation and surface region, pixels with missing or unphysical values, identified as non-finite values or values below $-10^{30}$, were masked prior to spectral extraction. 
The mean and median reflectance spectra were then computed over all valid pixels within the selected geographic region, using the same latitude and longitude boundaries defined in the mapping file. The resulting spectra were saved as two-column text files containing the CRISM wavelength array and the corresponding reflectance values. \\

The solar longitude criterion defined at the beginning of Sect.~\ref{subsec:dataretrival} was applied to the CRISM data as well, but its implementation differs from the OMEGA case because of the nature of the products. The VRDR tiles are mosaics assembled from several acquisitions obtained at different times and, consequently, at different seasons, so the criterion could not be applied to a product as a whole. In those cases the spectral extraction was restricted to the portions of the region covered by acquisitions falling within the adopted $L_s$ window, and the remaining pixels were masked before computing the mean and median spectra. The complete mosaics are shown in Fig.~\ref{fig:region_comparison} for context only and are not used in the derivation of the correction factors. In the subsequent analysis pipeline, CRISM spectral values within the HS-H wavelength range were retained.  However, the VNIR detector of CRISM carries a two-segment order-sorting filter whose joint lies near 0.65~$\mu$m, masked by an opaque stripe that attenuates the signal between approximately 0.61 and 0.71~$\mu$m \citep{Murchie2007}. The channels affected are not provided in the products used here, which leaves a gap between approximately 0.638 and 0.690~$\mu$m in the retrieved spectra. The first three HS-H bands (0.661, 0.670, and 0.688~$\mu$m) fall within this gap and were therefore excluded from the comparison with CRISM. A total of 47 CRISM observations, covering the same planetographic coordinates as the OMEGA observations described in Section~\ref{subsubsec:OMEGA_extraction}, were used in the analysis.

\subsection{Spectral Preprocessing}
\label{subsec:preprocessing}
To enable a direct spectral comparison between HS-H, OMEGA, and CRISM, the continuous spectra from OMEGA and CRISM must be projected onto the same 25 spectral bands as HS-H by applying the HS-H filter transfer functions (Eq.~\ref{eq:transfer}). For each spectral band $i$, the synthetic HS-H-equivalent value is obtained by integrating the product of the input spectrum $S(\lambda)$ and the filter transfer function $F^i(\lambda)$, weighted by $\lambda$ to account for the conversion between frequency and wavelength domains. Both mean and median spectra are processed independently through this procedure. \\

\begin{figure}[!ht]
\centering
\includegraphics[width=0.9\linewidth]{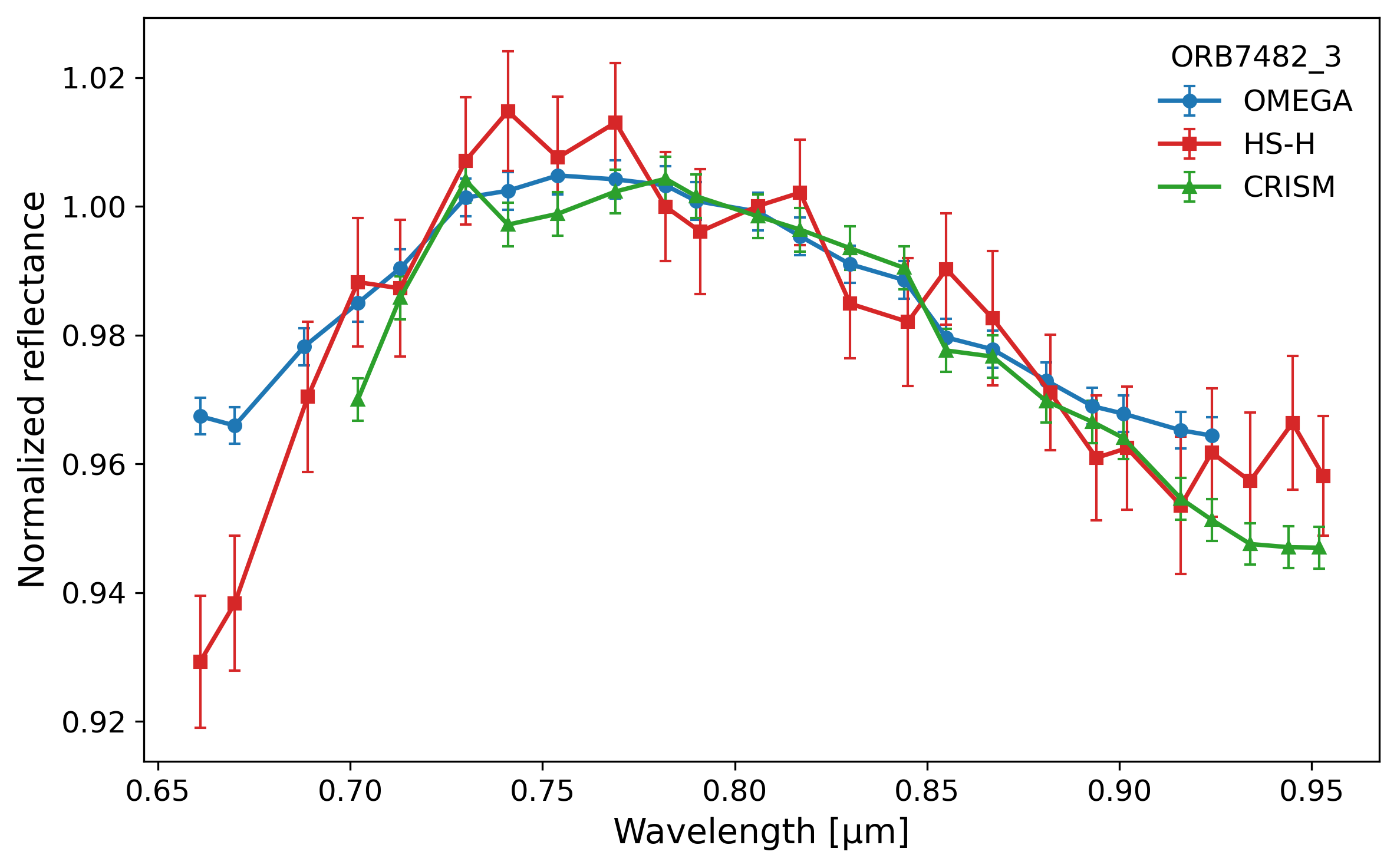}
\caption{Example of normalized spectra for the Huygens crater region ($ORB7482\_3$) as observed by HS-H (red), OMEGA (blue), and CRISM  (green), after application of the HS-H filter transfer functions and normalization at 0.80~$\mu$m. The error bars on the HS-H spectrum combine the photon and read-out noise of the 94 macropixels averaged over the region with the relative spectral calibration uncertainty, as described in Sect.~\ref{subsec:calibration}. The 1.3\% absolute radiometric uncertainty is common to all 25 channels and cancels in the normalization, and is therefore not shown. The error bars on the OMEGA and CRISM spectra are the band-to-band scatter of each spectrum about a smooth continuum.}
\label{fig:spectra_regionH}
\end{figure}

The preprocessing pipeline consists of three sequential steps. The normalization steps are applied identically to OMEGA, CRISM and HS-H, while the filter integration concerns only OMEGA and CRISM, since HS-H already samples the 25 bands natively. First, each spectrum is normalized at 0.80~$\mu$m, computed as the median of all spectral values within the window $[0.78, 0.82]$~$\mu$m, to remove the effect of the absolute flux level and place all spectra on a common relative scale. This step is performed prior to the filter integration. The normalized spectrum is then integrated using Eq.~\ref{eq:transfer} to produce a 25-band synthetic spectrum. Finally, the resulting spectrum is normalized again at 0.80~$\mu$m to remove any residual numerical offset introduced by the discrete integration. An example of the resulting normalized spectra for a representative surface region is shown in Fig.~\ref{fig:spectra_regionH}. \\

Prior to the preprocessing, a quality selection was applied to the OMEGA and CRISM spectra. Observations affected by saturation, low signal-to-noise ratio, or containing invalid values were discarded. Surface regions that fell outside the spatial coverage of a given HS-H image were also excluded.  In addition to these automated criteria, a visual inspection of all spectra was performed after the spectral extraction and normalization. Regions for which the HS-H spectrum showed insufficient signal-to-noise ratio, strong noise features, or incomplete spatial coverage within the image field of view were discarded. Additionally, regions with a heterogeneous mix of bright and dark terrain were excluded, as the spatial resolution mismatch between HS-H and OMEGA would render the spectral comparison unreliable in such cases.

\subsection{Observation selection}
\label{subsec:dataretrival}

The three HS-H images were acquired at a solar longitude of $L_s \simeq 55.7$\textdegree, that is, close to aphelion, during northern spring. At this season the atmospheric dust opacity reaches its annual minimum, whereas the period between $L_s = 180$\textdegree{} and $L_s = 360$\textdegree{} encompasses the dust storm season, during which the column opacity can increase by a large factor and occasionally culminates in planet-encircling events \citep{Montabone2015, SanchezLavega2024}. Because dust scattering raises the apparent reflectance at the shorter wavelengths of the HS-H range and flattens the spectral slope, comparing HS-H spectra against reference spectra acquired under a substantially higher dust load would introduce a seasonal bias that is indistinguishable from an instrumental one. For this reason, only OMEGA and CRISM observations acquired between $L_s = 0$\textdegree{} and $L_s = 180$\textdegree, that is, during northern spring and summer, were retained for the comparison. A second criterion was applied to the illumination geometry: observations whose OMEGA solar incidence angle was larger than 55\textdegree{} were discarded, since under grazing illumination the Lambertian photometric correction common to the three datasets becomes increasingly inaccurate and the surface is more sensitive to local topography. Both criteria were applied to the observation and not to the instrument, so that the OMEGA and the CRISM comparisons are defined by the same set of observing epochs and the two reference datasets sample identical surfaces under identical seasonal and illumination conditions. Out of the observations initially available, this selection retains 47 region--orbit pairs distributed over 21 surface regions and 26 distinct Mars Express orbits. The retained observations and their observing geometry are listed in Table~\ref{tab:geometry_appendix}, and the corresponding ranges of the photometric angles are summarized in Table~\ref{tab:angles}.

\FloatBarrier

\section{Results}
\label{sec:results}

\subsection{Spatial matching}

The key requirement for cross-calibrating HS-H, OMEGA, and CRISM is the definition of a common reference coordinate system for all observations. For images acquired over the Martian surface, we used Mars planetographic coordinates, expressed in latitude and longitude. The OMEGA and CRISM datasets already include spatial information referenced to Martian latitudes and longitudes.

To assign latitude and longitude coordinates to each HS-H pixel, we used the ESA Hera SPICE Kernel Dataset \citep{esa_hera_spice_2024}. SPICE, an acronym for Spacecraft, Planet, Instrument, C-matrix, and Events, is an information system that consists of data files, referred to as kernels, and software routines used to compute the geometric parameters of observations acquired by instruments onboard spacecraft. We therefore retrieved the latitude and longitude corresponding to the HS-H pointing geometry, including the camera center and image corners. The HS-H field distortion model \citep{Prodan2026} was then taken into account by applying the corresponding distortion corrections in order to assign latitude and longitude coordinates to each subpixel.

Prior to spectral extraction, the planetographic co-registration between the HS-H and OMEGA datasets was visually assessed for all 21 surface regions. For each region, the latitude-longitude bounding box defined for spectral extraction was overlaid on both the HS-H image and the corresponding OMEGA image. The spatial coincidence of the selected area was then confirmed by visual inspection. The comparison was guided by distinct morphological features, including impact craters, topographic variations, and other recognizable surface structures. These features served as spatial reference markers to assess the consistency of co-registration between the datasets. Across the examined regions, no visually identifiable discrepancies or spatial offsets were observed,  confirming that the co-registration is accurate at the scale of the regions used for the comparison.

\begin{table}[!ht]
\centering
\caption{Photometric angles of the OMEGA, CRISM and HS-H observations retained for the comparison, after the selection in solar longitude and incidence angle described in Sect.~\ref{subsec:dataretrival}. Values are the minimum and maximum over the 47 retained region--orbit pairs, with the mean in parentheses. All angles are in degrees.}
\label{tab:angles}
\begin{tabular}{lccc}
\hline\hline
Instrument & Phase & Emission & Incidence \\
\hline
OMEGA     & 17.0--54.6 (33.5) & 2.2--7.7 (4.3)   & 16.9--54.0 (33.1) \\
CRISM     & 41.1--62.1 (51.8) & 0.5--5.8 (1.5)   & 41.4--62.1 (52.1) \\
HS-H (04) & 19.8--27.2 (23.6) & 3.9--37.3 (20.7) & 6.8--44.2 (25.0) \\
HS-H (05) & 25.6--33.3 (29.6) & 5.7--39.8 (20.9) & 10.2--45.2 (27.5) \\
HS-H (06) & 34.6--41.0 (37.9) & 1.0--36.6 (19.4) & 22.1--46.3 (34.0) \\
\hline
\end{tabular}
\end{table}

\subsection{Comparison of normalized spectra}

\label{subsec:CF_norm}
The wavelength-dependent correction factors between HS-H and OMEGA, and between HS-H and CRISM, were derived from the normalized spectra described in Section~\ref{subsec:preprocessing}. For each surface region and each OMEGA observation, the correction factor at band $i$ is defined as the ratio of the normalized HS-H spectrum to the normalized OMEGA (or CRISM) spectrum after application of the HS-H filter transfer functions:

\begin{equation}
    C^i = \frac{S^i_\mathrm{HH,\,norm}}{S^i_\mathrm{OMEGA,\,norm}}
    \label{eq:CF}
\end{equation}

where $S^i_\mathrm{HH,\,norm}$ and $S^i_\mathrm{OMEGA,\,norm}$ are the normalized reflectance values at band $i$ for HS-H and OMEGA respectively. An equivalent definition applies for CRISM.  The correction factors were computed independently from the mean and the median spectra; since the two agree to better than 0.2\% across all regions and bands, only the median ones are presented hereafter, as they are more robust to outliers. \\

For each band, the correction factor was computed separately for each OMEGA observation and surface region. The median and standard deviation across all observations and regions were then computed to obtain a robust estimate of the systematic offset between the instruments. This procedure was applied independently to each of the three HS-H images (HS-L0-FE-025D-04-01, HS-L0-FE-025D-05-01, and HS-L0-FE-025D-06-01), referred to hereafter as images~04, 05, and~06. The final correction factors were obtained by taking the median of the three per-image estimates, with the uncertainty computed as the maximum between the standard deviation among the three images and the median of the per-image uncertainties, ensuring a conservative error estimate.\\

\begin{figure}[!ht]
\centering

\begin{subfigure}{\linewidth}
    \centering
    \includegraphics[width=0.9\linewidth]{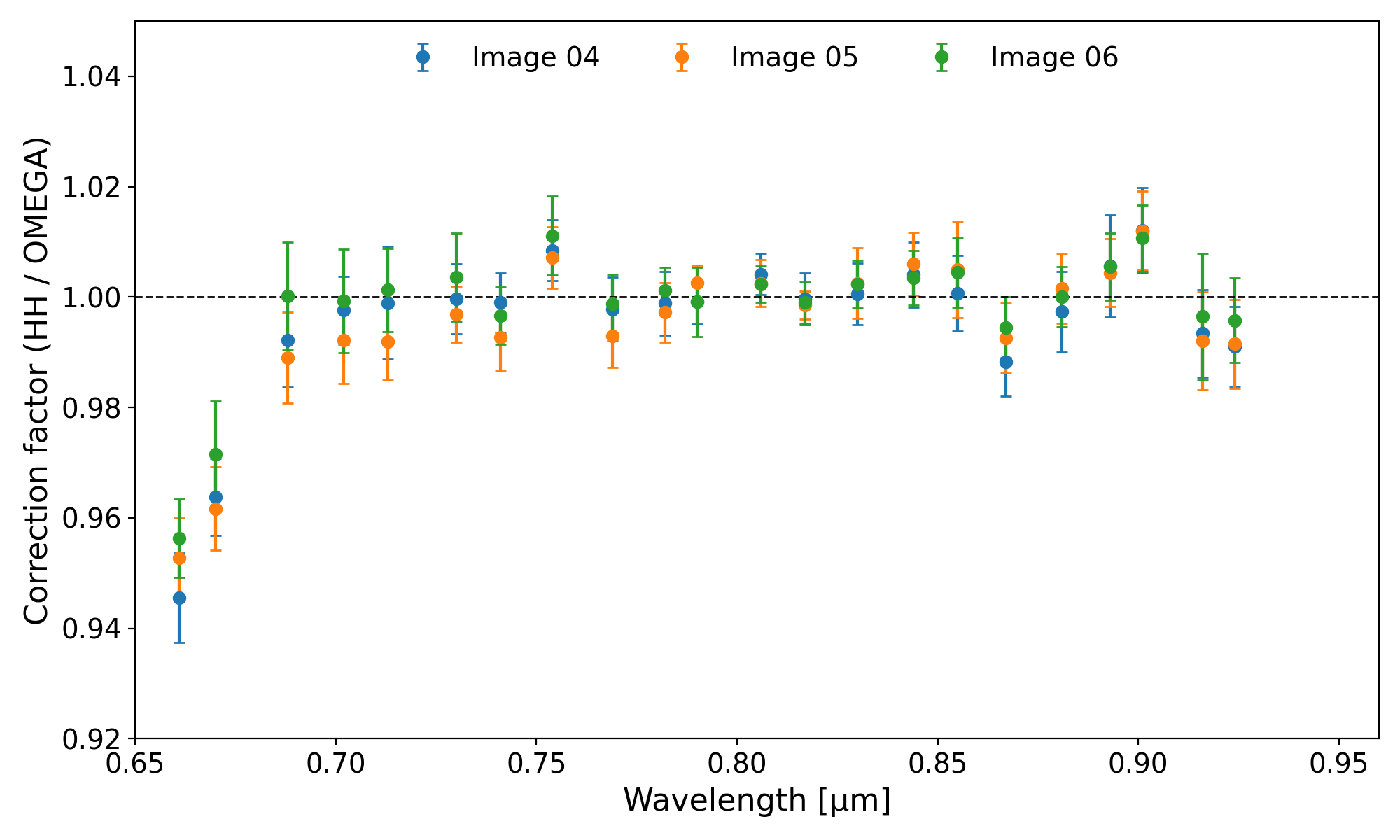}
\end{subfigure}

\vspace{0.3cm}

\begin{subfigure}{\linewidth}
    \centering
    \includegraphics[width=0.9\linewidth]{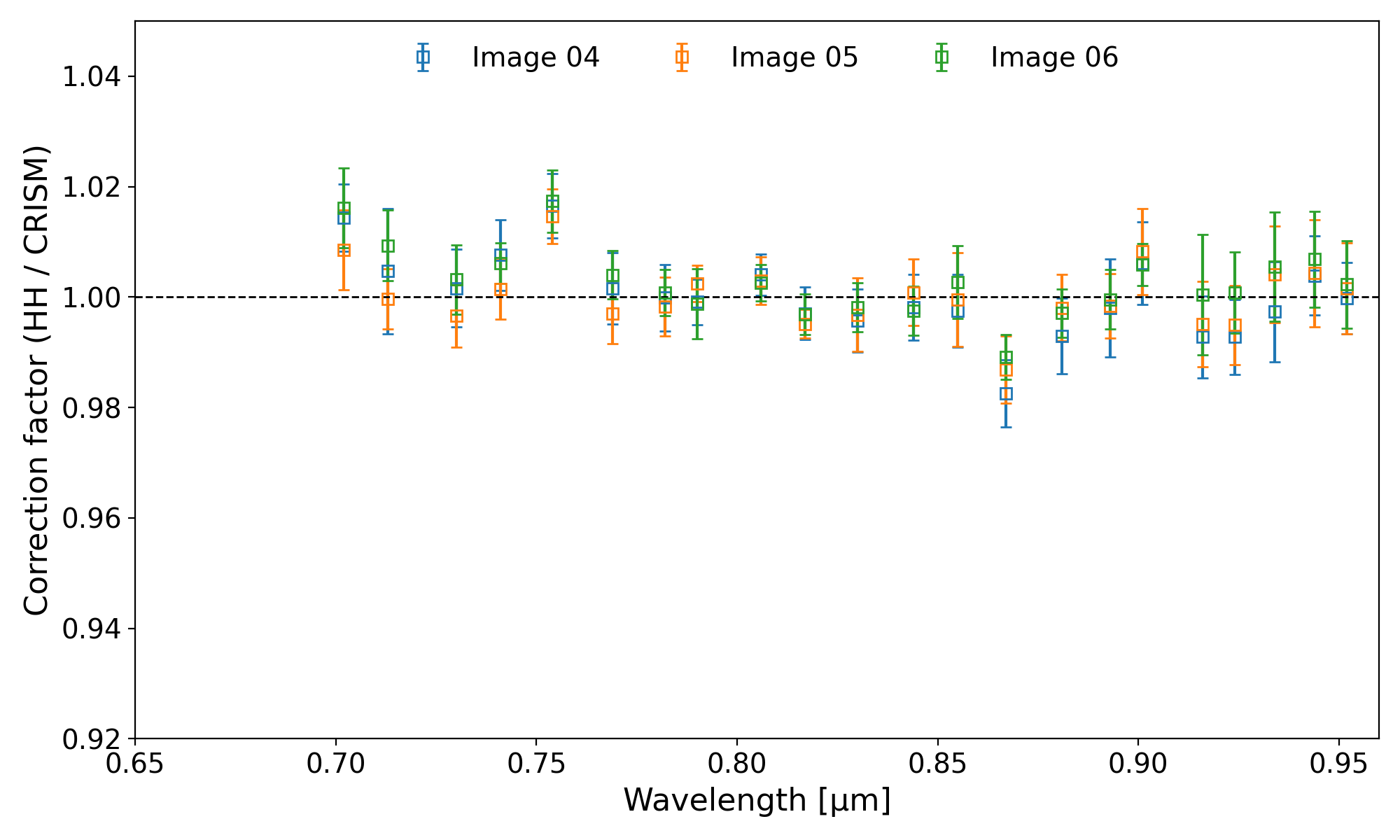}
\end{subfigure}

\caption{Wavelength-dependent correction factors $C^i$ between HS-H and OMEGA (upper panel) and between HS-H and CRISM (lower panel), derived from normalized 
spectra over the 21 selected surface regions. Blue, orange, and green points correspond to HS-H images ~04, ~05, and ~06, respectively. Error bars represent the 
standard deviation of the correction factors across all surface regions. The horizontal dashed line at unity indicates perfect agreement between the instruments.}

\label{fig:CF_norm}
\end{figure}

The resulting correction factors are shown in Fig.~\ref{fig:CF_norm} and listed in Table~\ref{tab:CF_norm}.  \\
For HS-H/OMEGA, the correction factors are close to unity for most bands.  The relative outliers are bands~1 and~2, with values of $0.9527 \pm 0.0072$ and $0.9639 \pm 0.0075$ at 0.661 and 0.670~$\mu$m, respectively, indicating a systematic offset of 4.7 and 3.6\% in the two shortest-wavelength bands. From band~3 (0.688~$\mu$m) onwards, the correction factors are close to unity, with a standard deviation below 1\% across all bands. Over bands~3--22, the median correction factor is  $0.9991 \pm 0.0052$. \\

For HS-H/CRISM,  the correction factors from band~4 (0.702~$\mu$m) onwards are  close to unity, with a median correction factor of  $0.9995 \pm 0.0066$ over bands~4--22. A few channels, in particular bands~8, 17 and~20, depart from unity by up to 1.7\%, with the same sign in the OMEGA comparison. Such residuals may originate either from the relative spectral calibration of HS-H or from differences between the two reference datasets, and are precisely what the wavelength-dependent correction factors are meant to account for. Band~4 is a separate case, since it lies at the edge of the CRISM spectral gap between 0.638 and 0.690~$\mu$m, where the HS-H transfer functions are partly evaluated by extrapolation.

\begin{table}[htbp]
\centering
\caption{normalized correction factors of HS-H with respect to MEx/OMEGA and MRO/CRISM,
derived from the orbits with $L_s$ between 0\textdegree{} and 180\textdegree{} and solar
incidence angle below 55\textdegree. Values are the median over the three HS-H images
(04, 05, 06); the uncertainty is the maximum between the dispersion among the three images and the median of the per-image dispersions across the surface regions. In the last row the median is taken over the bands listed in the corresponding column and the uncertainty is their dispersion. The first three bands are not
covered by CRISM and the last three are not covered by OMEGA.}
\label{tab:CF_norm}
\begin{tabular}{c c c c}
\hline
Band & $\lambda$ [\textmu m] & CF (HS-H / OMEGA) & CF (HS-H / CRISM) \\
\hline
1  & 0.661 & $0.9527 \pm 0.0072$ & --- \\
2  & 0.670 & $0.9639 \pm 0.0075$ & --- \\
3  & 0.688 & $0.9922 \pm 0.0086$ & --- \\
4  & 0.702 & $0.9976 \pm 0.0079$ & $1.0144 \pm 0.0072$ \\
5  & 0.713 & $0.9990 \pm 0.0075$ & $1.0047 \pm 0.0064$ \\
6  & 0.730 & $0.9996 \pm 0.0063$ & $1.0016 \pm 0.0063$ \\
7  & 0.741 & $0.9966 \pm 0.0054$ & $1.0061 \pm 0.0054$ \\
8  & 0.754 & $1.0084 \pm 0.0056$ & $1.0165 \pm 0.0056$ \\
9  & 0.769 & $0.9978 \pm 0.0057$ & $1.0015 \pm 0.0055$ \\
10 & 0.782 & $0.9989 \pm 0.0054$ & $0.9999 \pm 0.0053$ \\
11 & 0.790 & $0.9992 \pm 0.0040$ & $0.9991 \pm 0.0041$ \\
12 & 0.806 & $1.0025 \pm 0.0038$ & $1.0030 \pm 0.0037$ \\
13 & 0.817 & $0.9990 \pm 0.0037$ & $0.9969 \pm 0.0037$ \\
14 & 0.830 & $1.0023 \pm 0.0056$ & $0.9968 \pm 0.0057$ \\
15 & 0.844 & $1.0041 \pm 0.0057$ & $0.9982 \pm 0.0060$ \\
16 & 0.855 & $1.0044 \pm 0.0068$ & $0.9995 \pm 0.0066$ \\
17 & 0.867 & $0.9926 \pm 0.0062$ & $0.9868 \pm 0.0060$ \\
18 & 0.881 & $1.0000 \pm 0.0062$ & $0.9971 \pm 0.0060$ \\
19 & 0.893 & $1.0055 \pm 0.0062$ & $0.9984 \pm 0.0058$ \\
20 & 0.901 & $1.0120 \pm 0.0072$ & $1.0061 \pm 0.0075$ \\
21 & 0.916 & $0.9934 \pm 0.0089$ & $0.9951 \pm 0.0077$ \\
22 & 0.924 & $0.9915 \pm 0.0077$ & $0.9949 \pm 0.0072$ \\
23 & 0.934 & ---                & $1.0041 \pm 0.0090$ \\
24 & 0.944 & ---                & $1.0044 \pm 0.0087$ \\
25 & 0.952 & ---                & $1.0016 \pm 0.0080$ \\
\hline
Median & & $0.9990 \pm 0.0131$ & $1.0007 \pm 0.0062$ \\
\hline
\end{tabular}
\end{table}

\subsection{The radiometric correction factor}
\label{subsec:CF_flux}

The correction factors described in Section~\ref{subsec:CF_norm} were derived from normalized spectra and therefore capture only the wavelength-dependent shape of the instrumental responses, removing any information on the absolute reflectance level. The radiometric correction factors corresponding to each channel are obtained directly from the relative correction factors derived in Section~\ref{subsec:CF_norm} by multiplying them by the constant (called the radiometric correction factor) defined as the ratio of the HS-H to OMEGA (or CRISM) reflectance at 0.80~$\mu$m (the same wavelength at which the normalization was performed previously).
This radiometric correction factor is  $0.9209 \pm 0.0095$ for HS-H/OMEGA and  $0.9933 \pm 0.0160$ for HS-H/CRISM, computed over bands~3--22 and bands~4--22 respectively. \\

All three instruments — OMEGA, CRISM, and HS-H — reduce their data to reflectance factor (I/F divided by the cosine of the incidence angle) following equivalent photometric conventions \citep{Mustard2007, Viviano2014}. The absolute flux differences observed between the instruments are therefore not solely attributable to differences in photometric conventions, but may reflect a combination of factors including residual differences in the absolute radiometric calibration of the instruments, differences in how the cosine of the incidence angle is computed with respect to the reference surface, and observation geometry effects arising from the different phase, emission, and incidence angles at which each instrument observed the same surface region. \\

\begin{figure}[!ht]
\centering

\begin{subfigure}{\linewidth}
    \centering
    \includegraphics[width=0.9\linewidth]{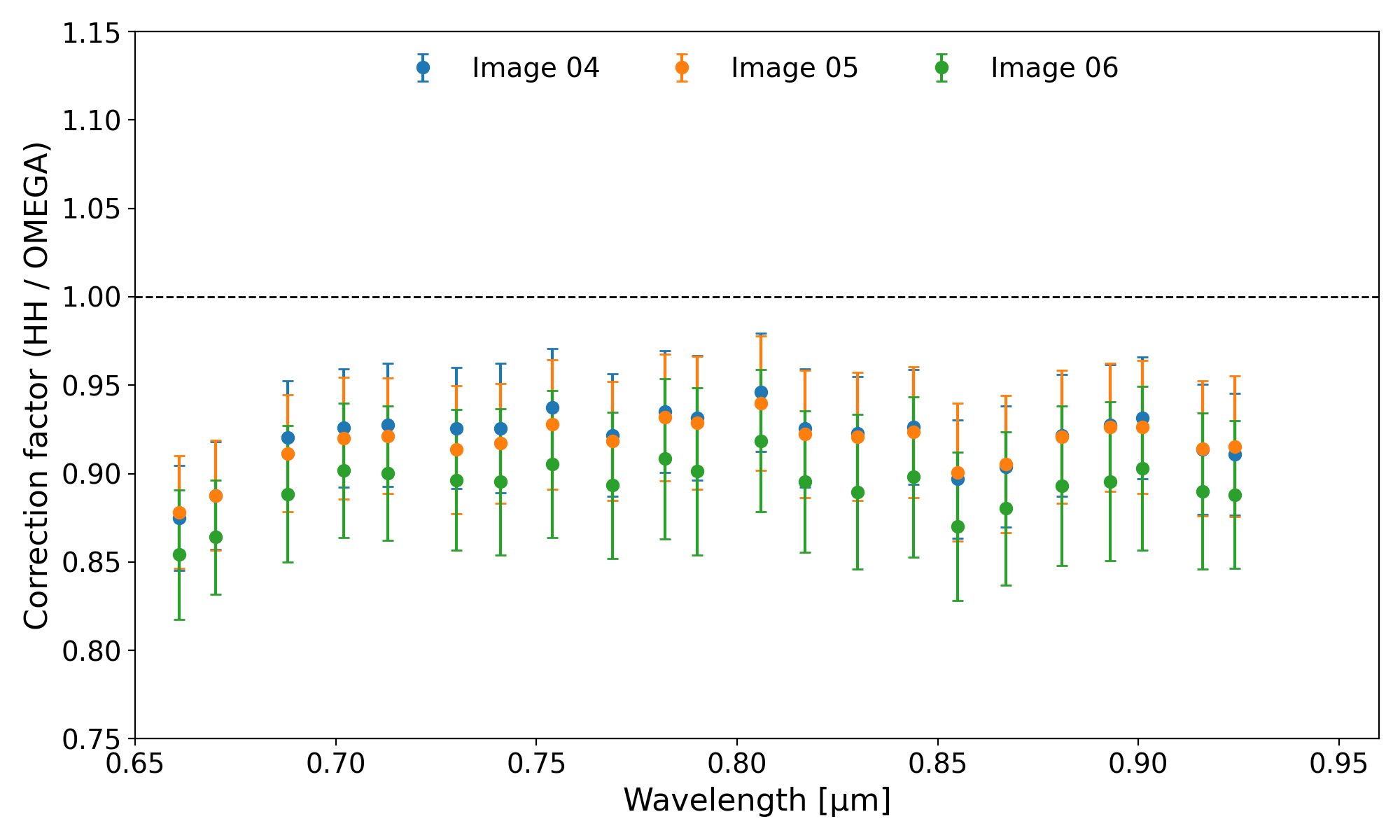}
\end{subfigure}

\vspace{0.3cm}

\begin{subfigure}{\linewidth}
    \centering
    \includegraphics[width=0.9\linewidth]{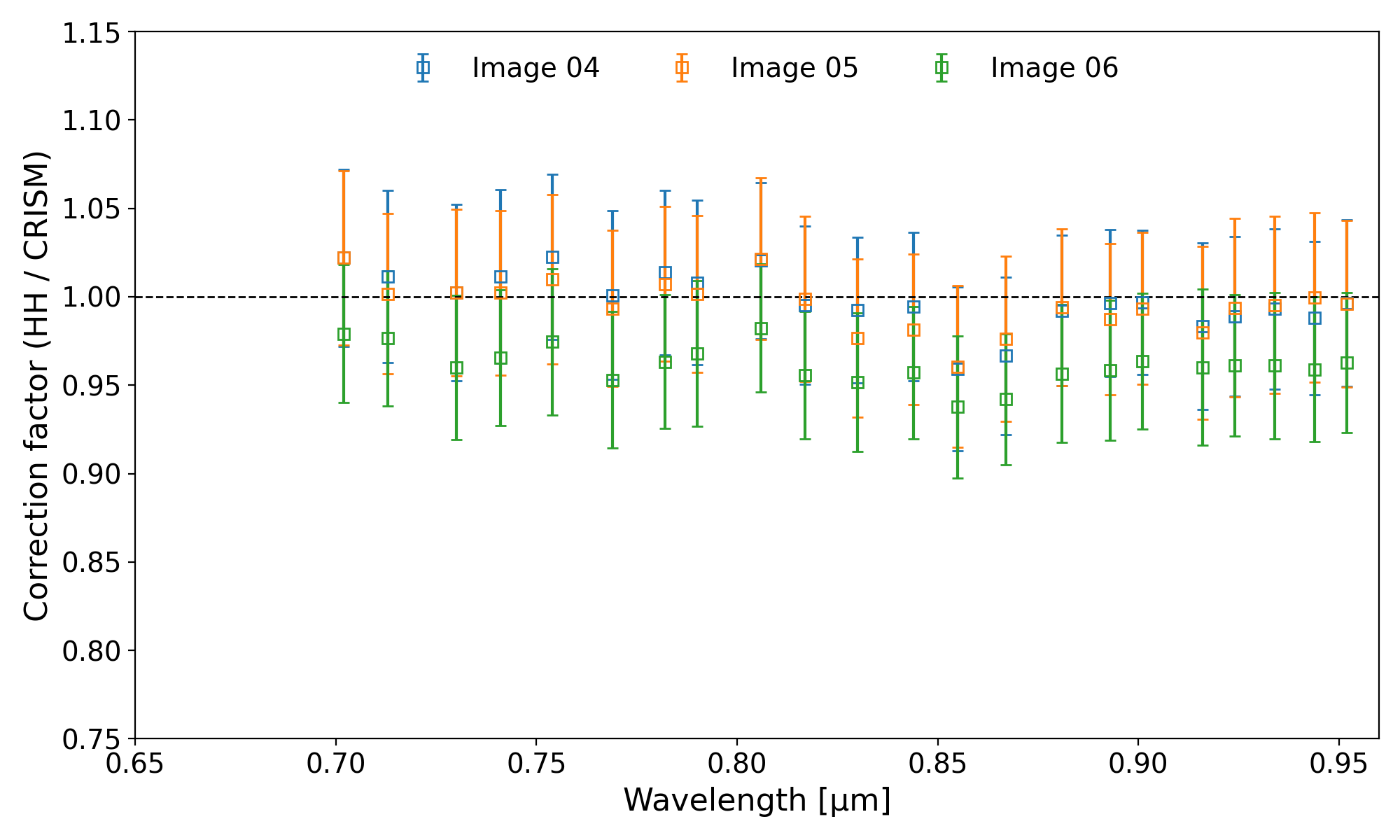}
\end{subfigure}

\caption{Wavelength-dependent absolute correction factors $C^i_\mathrm{abs}$ between HS-H and OMEGA (upper panel) and between HS-H and CRISM (lower panel). Blue, orange, and green points correspond to HS-H images~04, 05, and~06, respectively. Error bars represent the standard deviation across all surface regions. The horizontal dashed line at unity indicates perfect agreement.}
\label{fig:CF_flux}
\end{figure}

The resulting radiometric correction factors (multiplication between the radiometric correction factor and the correction factors derived for normalized spectra) are shown in Fig.~\ref{fig:CF_flux}.  The three images are consistent with each other within their uncertainties across all bands, confirming the repeatability of the measurement.

For HS-H/OMEGA, the correction factors range between  $0.8968 \pm 0.0390$ at 0.855~$\mu$m and $0.9397 \pm 0.0379$ at 0.806~$\mu$m, with a median value of $0.9209 \pm 0.0095$ across bands~3--22. This indicates that HS-H measures systematically lower absolute fluxes than OMEGA by approximately  8\%.  The per-band uncertainty of about 0.04 reflects the variation in surface brightness across the 21 selected regions. For HS-H/CRISM, the correction factors range between  $0.9594 \pm 0.0459$ at 0.855~$\mu$m and $1.0219 \pm 0.0492$ at 0.702~$\mu$m, with a median value of $0.9933 \pm 0.0160$ across bands~4--22. This indicates that HS-H and CRISM are in very good agreement in absolute flux, with a residual offset of less than 1\%.  The corresponding uncertainty, about 0.045, is slightly larger than for the OMEGA comparison. Both offsets are practically wavelength-independent: the band-to-band variation of the radiometric correction factors is comparable to the uncertainty of a single band, and the normalized correction factors of Table~\ref{tab:CF_norm}, which are several times more precise, vary by only 2 to 3\% over the same bands. This indicates that the radiometric discrepancy is multiplicative and does not distort the spectral shape, a favorable property for future scientific applications. \\

\begin{figure*}[!t]
    \centering
    \includegraphics[width=\linewidth]{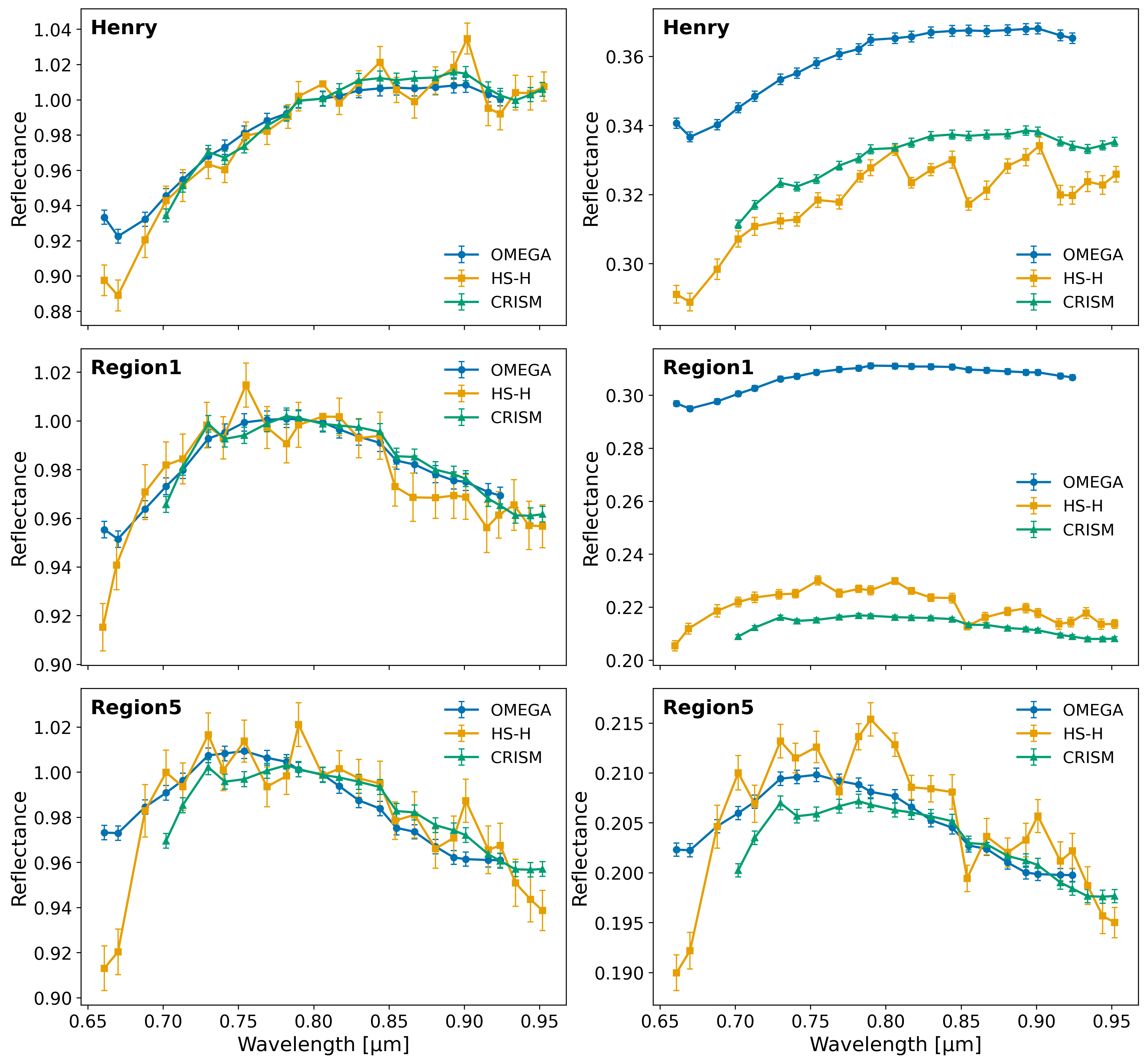}
    \caption {Normalized (left column) and absolute flux (right column) spectra for three representative surface regions — Henry crater (top), Region~1  (middle), and Region~5  (bottom) — as observed by OMEGA (blue), HS-H (orange), and CRISM (green). In the normalized case, all spectra are divided by their value at 0.80~$\mu$m. The error bars on the HS-H spectra combine the photon and read-out noise of the macropixels averaged over each region (104, 99 and 92 for Henry, Region~1 and Region~5 respectively) with the relative spectral calibration uncertainty, as described in Sect.~\ref{subsec:calibration}. The 1.3\% absolute radiometric uncertainty is common to all 25 channels: it cancels in the normalized panels and acts as a common scale factor in the absolute flux panels, and is therefore not shown. The error bars on the OMEGA and CRISM spectra are the band-to-band scatter of each spectrum about a smooth continuum.}
    \label{fig:spectra_regionX}
\end{figure*}

The representative spectra shown in Fig.~\ref{fig:spectra_regionX} illustrate the typical behavior of the three instruments for individual regions in both relative and absolute flux. In the normalized case, the three instruments show good agreement in spectral shape, confirming that the wavelength-dependent correction factors derived in Section~\ref{subsec:CF_norm} are close to unity. In absolute flux, HS-H is in good agreement with CRISM while measuring systematically lower fluxes than OMEGA, consistent with the correction factors reported here. Notably, OMEGA and CRISM do not always agree in absolute flux — a behavior that varies from region to region and is consistent with the different observation geometries of the two instruments over the same surface area. The analysis of the dependence of the absolute correction factors on observation geometry is presented in Section~\ref{subsec:geometry}.

\begin{table}[htbp]
\centering
\caption{Absolute-flux correction factors of HS-H with respect to MEx/OMEGA and MRO/CRISM,
derived from the same orbit selection as Table~\ref{tab:CF_norm}. Values are the median over
the three HS-H images (04, 05, 06); the uncertainty is the maximum between the dispersion among the three images and the median of the per-image dispersions across the surface regions. In the last row the median is taken over the bands listed in the corresponding column and the uncertainty is their dispersion. The first
three bands are not covered by CRISM and the last three are not covered by OMEGA.}
\label{tab:CF_flux}
\begin{tabular}{c c c c}
\hline
Band & $\lambda$ [\textmu m] & CF (HS-H / OMEGA) & CF (HS-H / CRISM) \\
\hline
1  & 0.661 & $0.8748 \pm 0.0319$ & --- \\
2  & 0.670 & $0.8877 \pm 0.0311$ & --- \\
3  & 0.688 & $0.9114 \pm 0.0330$ & --- \\
4  & 0.702 & $0.9198 \pm 0.0345$ & $1.0219 \pm 0.0492$ \\
5  & 0.713 & $0.9213 \pm 0.0348$ & $1.0016 \pm 0.0453$ \\
6  & 0.730 & $0.9135 \pm 0.0363$ & $1.0023 \pm 0.0470$ \\
7  & 0.741 & $0.9170 \pm 0.0365$ & $1.0022 \pm 0.0464$ \\
8  & 0.754 & $0.9278 \pm 0.0367$ & $1.0099 \pm 0.0466$ \\
9  & 0.769 & $0.9184 \pm 0.0347$ & $0.9933 \pm 0.0442$ \\
10 & 0.782 & $0.9317 \pm 0.0358$ & $1.0071 \pm 0.0437$ \\
11 & 0.790 & $0.9288 \pm 0.0376$ & $1.0016 \pm 0.0442$ \\
12 & 0.806 & $0.9397 \pm 0.0379$ & $1.0204 \pm 0.0441$ \\
13 & 0.817 & $0.9223 \pm 0.0361$ & $0.9951 \pm 0.0447$ \\
14 & 0.830 & $0.9209 \pm 0.0362$ & $0.9765 \pm 0.0413$ \\
15 & 0.844 & $0.9234 \pm 0.0370$ & $0.9815 \pm 0.0421$ \\
16 & 0.855 & $0.8968 \pm 0.0390$ & $0.9594 \pm 0.0459$ \\
17 & 0.867 & $0.9039 \pm 0.0388$ & $0.9665 \pm 0.0445$ \\
18 & 0.881 & $0.9209 \pm 0.0375$ & $0.9922 \pm 0.0426$ \\
19 & 0.893 & $0.9262 \pm 0.0361$ & $0.9873 \pm 0.0416$ \\
20 & 0.901 & $0.9263 \pm 0.0377$ & $0.9933 \pm 0.0409$ \\
21 & 0.916 & $0.9138 \pm 0.0382$ & $0.9797 \pm 0.0472$ \\
22 & 0.924 & $0.9109 \pm 0.0399$ & $0.9888 \pm 0.0450$ \\
23 & 0.934 & ---                & $0.9931 \pm 0.0454$ \\
24 & 0.944 & ---                & $0.9879 \pm 0.0432$ \\
25 & 0.952 & ---                & $0.9959 \pm 0.0471$ \\
\hline
Median & & $0.9203 \pm 0.0144$ & $0.9933 \pm 0.0149$ \\
\hline
\end{tabular}
\end{table}

\FloatBarrier

\section{Discussion}
\label{sec:Discussion}
\subsection{Repeatability and Reliability of the measurements}
\label{subsec:repeatability}

The three HS-H images used in this work (04, 05, and~06) were acquired over the same surface area at different distances from Mars (20924.8, 17610.2, and 14521.4~km respectively) and with slightly different viewing geometries. The availability of three independent observations of the same surface regions allows a direct assessment of measurement repeatability and of the reliability. \\

For the relative correction factors, the three HS-H images produce highly consistent results across all 25 spectral bands, as shown on Fig.~\ref{fig:CF_norm}. The inter-image standard deviation, computed band by band across the three estimates, has a mean value of  0.22\% for HS-H/OMEGA and  0.20\% for HS-H/CRISM. The maximum inter-image standard deviation reaches  0.47\% at 0.688~$\mu$m for OMEGA and  0.39\% at 0.713~$\mu$m for CRISM, with maximum point-to-point differences between any two images of  1.13\% and 0.96\%, respectively. These values are  comparable to the per-band
uncertainties reported in Table~\ref{tab:CF_norm}, confirming that the relative correction factors are highly reproducible across different observations of the same surface. \\

For the radiometric correction factor, inter-image consistency remains good, though the spread is larger due to the greater sensitivity of the absolute flux to variations in observation geometry. The mean inter-image standard deviation is  1.37\% for HS-H/OMEGA and  1.70\% for HS-H/CRISM, with maximum values of  1.65\% at 0.830~$\mu$m for OMEGA and  2.22\% at 0.782~$\mu$m for CRISM. The maximum point-to-point difference between any two images is  3.59\% for OMEGA and  5.01\% for CRISM, again  comparable to the per-band
uncertainties. These results confirm that the radiometric correction factor is stable across the three HS-H images despite different distances and viewing geometries, with the larger spread relative to the normalized case reflecting the sensitivity of the absolute flux to varying illumination geometry. \\

The reliability of the cross-calibration result is further supported by the large number of independent surface regions and OMEGA observations used (21 regions and 47  region--orbit pairs from 26 unique Mars Express orbits), which sample a wide variety of surface types and illumination conditions. The low inter-image standard deviation, combined with the consistency between the mean and median correction factors noted in Section~\ref{subsec:CF_norm}, demonstrates that the derived correction factors are robust against the choice of individual observations or surface regions. Absolute radiometric offsets between independently calibrated instruments are a well-known result in planetary remote sensing: an offset of 17--18\% was found between MapCam and OVIRS onboard OSIRIS-REx \citep{Golish2022}, and up to 15\% between ONC-T and MapCam \citep{Yumoto2024}, both attributed to differences in radiometric calibration procedures. The offset found here for HS-H with respect to OMEGA ( $\sim$8\%) is consistent with this level of inter-instrument variability. The close agreement between HS-H and CRISM ($<$1\%) is consistent with the use of CRISM observations to determine the relative gain factors of HS-H during the in-flight calibration \citep{Prodan2026}. \\ 

\subsection{Analysis of Observation Geometry}
\label{subsec:geometry}

\begin{figure*}[!ht]
    \centering
    \includegraphics[width=\linewidth]{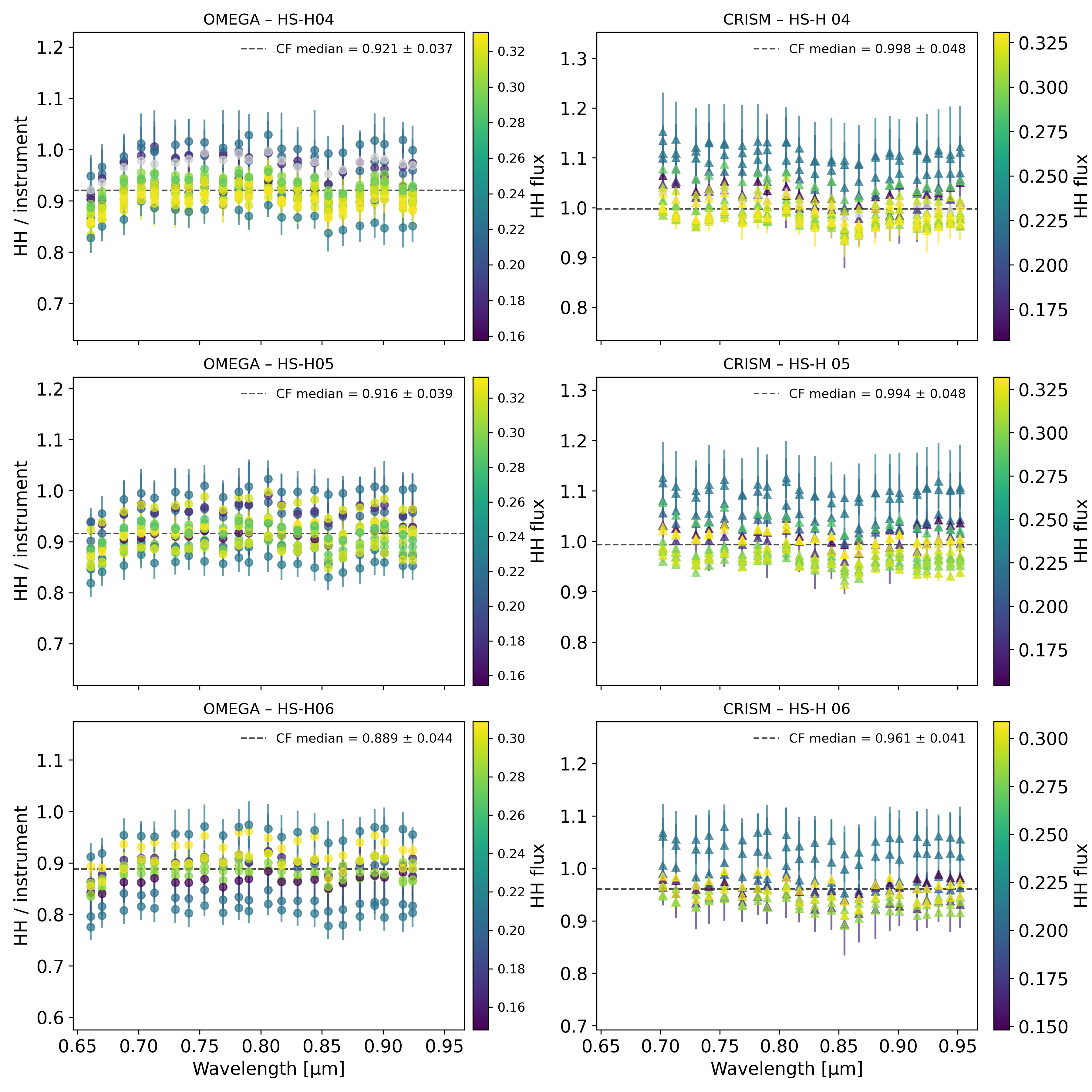}
    \caption{Radiometric correction factors $C^i_\mathrm{abs}$ for all surface regions retained after quality selection, shown as a function of wavelength for each of the three HS-H images (rows, from top to bottom, images~04, 05, and~06) and for the two reference instruments (columns: OMEGA on the left, CRISM on the right). Each circle (OMEGA) or triangle (CRISM) represents the median correction factor of one surface region, computed over all available observations of that region for the given HS-H image. Error bars represent the standard deviation of the correction factor across the different observations of the same region. The color of each symbol indicates the median HS-H flux of the corresponding surface region, used as a proxy for surface albedo, from dark purple for the darkest regions to yellow for the brightest. The correction factors span a wide range of values across the selected regions, reflecting the combined influence of surface albedo and observing geometry on the absolute radiometric comparison between HS-H and the reference instruments.}
    \label{fig:flux_panel}
\end{figure*}

The radiometric correction factor shows significant dispersion across the different surface regions and observations. Figure~\ref{fig:flux_panel} shows the radiometric correction factors for the surface regions retained after quality selection, separately for the three HS-H images (04, 05, and 06) and the two reference instruments (OMEGA on the left, CRISM on the right). For each region, the correction factor shown is the median computed over all available OMEGA observations for that region, with error bars representing the standard deviation across the different observations. The color of each point represents the median HS-H flux for that region, computed as the median over all spectral bands and all available observations for that image — from dark purple for the darkest regions to yellow for the brightest. Circles indicate the comparison with OMEGA and triangles with CRISM. The panel shows that the correction factor varies significantly across the selected regions, with values ranging from approximately  0.78 to 1.03 for OMEGA and from approximately 0.90 to 1.15 for CRISM. This spread is substantially larger than the inter-image standard deviation discussed in Section~\ref{subsec:repeatability}, indicating that the region-to-region variability cannot be attributed to instrumental noise alone. \\

The photometric correction applied to the HS-H data consists of a Lambertian model based on division by the cosine of the local incidence angle. The Martian surface is known to deviate from Lambertian behavior, with Minnaert exponents $k$ ranging from approximately 0.55 for the darkest regions to 0.95 for the brightest ones \citep{Erard2001, Munaretto2021}, which may contribute to the observed dispersion. The two reference instruments show a mutual offset of approximately  7.3\%, with OMEGA measuring systematically higher fluxes than CRISM across the entire spectrum ( CRISM/OMEGA ratio $= 0.927 \pm 0.010$ over bands~4--22). This offset is consistent with differences in the absolute radiometric calibrations of the two instruments, which were performed independently using different procedures and calibration targets \citep{Bibring2004, Murchie2007}. The close agreement between HS-H and CRISM is consistent with the in-flight calibration procedure described in \citet{Prodan2026}, in which the relative gain factors of HS-H across spectral channels were determined using CRISM observations, while the absolute sensitivity was independently derived from stellar observations. The offset between HS-H and OMEGA may therefore reflect the same calibration differences that produce the mutual offset between OMEGA and CRISM. Offsets of similar magnitude have been reported in the literature for other planetary mission instruments. \citet{Golish2022} and \citet{Yumoto2024} found offsets of up to 15--18\% between independently calibrated instruments, confirming that this level of inter-instrument variability is expected in planetary remote sensing. \\

\subsection{The behavior of the first two HS-H bands}
\label{sec:bands}
The correction factors derived from the normalized spectra exhibit distinct behavior between the first two bands and the remainder of the spectrum. For bands 3--25 (0.688--0.952~$\mu$m), that is bands 3--22 for OMEGA and bands 4--25 for CRISM, all relative correction factors are close to  unity, with a standard deviation below 1\% for both the OMEGA and CRISM comparisons. This indicates that the relative spectral response of HS-H is in good agreement with that of the reference instruments across the entire operational wavelength range of the instrument, and that the HS-H filter transfer functions correctly describe the in-flight spectral shape of the instrument. Bands 1 and 2 (0.661 and 0.670~$\mu$m) instead show sub-unity values systematically, with  $C^i = 0.9527 \pm 0.0072$ and $C^i = 0.9639 \pm 0.0075$, respectively, for comparison with OMEGA. These values indicate that HS-H measures approximately  4.7\% and 3.6\% less than OMEGA at the shortest wavelengths. This offset is absent in the subsequent bands,  and must be related to an instrumental effect confined to the two shortest bands. For this reason, bands 1 and 2 were excluded  when computing the median radiometric correction factor, and their careful assessment is recommended for future scientific analyzes of HS-H spectra. \\

\begin{figure}[!ht]
\centering
\includegraphics[width=0.9\linewidth]{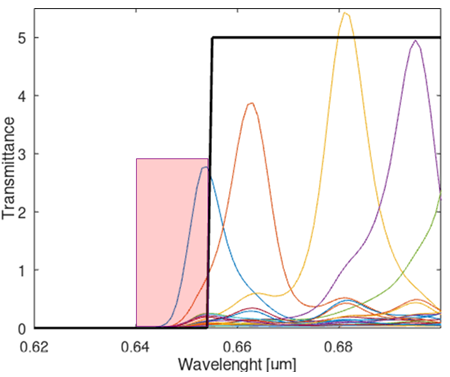}
\caption{Illustration of the wavelength drift effect on the HS-H filter transfer functions. The right panel shows a zoom on the two shortest bands (0.661 and 0.670~$\mu$m), where the primary peaks fall close to the blocking region of the broadband filter (red shaded area). At in-flight operating temperatures, the Fabry-Pérot plate spacing decreases, causing a shift of the transfer functions towards shorter wavelengths (blue arrow), which partially overlaps the blocking region and reduces the total transmission of these bands.}
\label{fig:popescu_drift}
\end{figure}

The cause of this offset was hypothesized to be a wavelength drift in the HS-H filter transfer functions, arising from temperature differences between the ground calibration and the in-flight operating conditions. The HS-H instrument uses a Fabry-Pérot filter for spectral separation, whose response exhibits multiple peaks at different wavelengths. A broadband blocking filter is included to suppress unwanted higher-order peaks, allowing only the primary peak of each band to pass through. For bands 1 and 2, the primary peak falls close to the edge of the broadband blocking filter, a configuration that makes these bands particularly sensitive to any variation in the spacing of the Fabry-Pérot plates. At temperatures lower than those used during ground calibration, the plate spacing decreases, causing a shift of the transmitted wavelengths towards shorter values. This shift moves the primary peaks of bands 1 and 2 closer to the blocking region of the broadband filter, drastically reducing the total transmission of these two bands and producing the signal deficit observed in the correction factors (Fig.~\ref{fig:popescu_drift}).

\begin{figure}[!ht]
\centering
\includegraphics[width=0.9\linewidth]{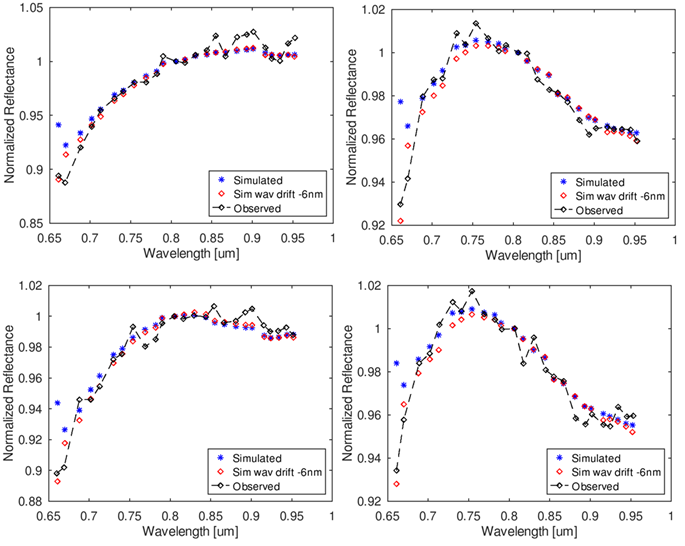}
\caption{Comparison of normalized spectra for four representative surface regions. Blue points: OMEGA spectra transformed using the nominal HS-H transfer functions. Red diamonds: OMEGA spectra transformed using the HS-H transfer functions shifted by $-6$~nm towards shorter wavelengths. Black diamonds: observed HS-H spectra for the same regions. The shift of $-6$~nm produces a significantly better agreement with the observed HS-H spectra at the two shortest wavelength bands, while differences for bands 3--25 are negligible and fall within the measurement uncertainties. Note that all data points are plotted using the standard (nominal) central wavelengths.}
\label{fig:popescu_spectra}
\end{figure}

To quantify this effect, the HS-H transfer functions were applied to OMEGA spectra of the same surface regions analyzed in this work, first using the nominal transfer functions and then using transfer functions shifted by different amounts until the best agreement with the observed HS-H spectra was found. The shift producing the best agreement is $-6$~nm, consistent with the temperature difference expected between the ground calibration (performed at room temperature) and the in-flight operating conditions during the Mars flyby (approximately $-12^{\circ}$C), which is physically plausible as the cause of the observed shift. As shown in Fig.~\ref{fig:popescu_spectra}, the OMEGA spectra transformed using the $-6$~nm shifted transfer functions (red curve) match the observed HS-H spectra (black curve) much better than those transformed with the nominal transfer functions (blue curve), particularly for bands 1 and 2. For bands 3--25, the difference between the two transfer functions is negligible and falls within the uncertainties, confirming that the drift is confined to the shortest wavelengths and does not affect the rest of the spectrum.  Band~4, the first band covered by CRISM, shows a value slightly above unity ($C^i = 1.0144 \pm 0.0072$) that is not attributable to the drift affecting bands 1 and 2. It lies at the edge of the CRISM spectral gap between 0.638 and 0.690~$\mu$m, where the HS-H transfer functions are partly evaluated by extrapolation, as discussed in Sect.~\ref{subsec:CF_norm}. \\

\FloatBarrier

\section{Conclusions}
\label{sec:Conclusions}

We have presented a comprehensive cross-calibration of the HyperScout-H, the hyperspectral imager onboard ESA's Hera spacecraft, using data acquired during the 12 March 2025 Mars flyby. The in-flight radiometric calibration of HS-H has followed a progressive path: the  ground-based calibration performed at ESTEC \citep{Popescu2025} established the baseline instrument response under laboratory conditions; the  in-flight calibration \citep{Prodan2026} refined the sensitivity coefficients and geometric distortion correction using stellar observations and a preliminary comparison with CRISM data over five Martian surface regions, providing the first in-flight estimate of the instrument performance. Building on this foundation, the present work constitutes the  cross-calibration step, extending the analysis to 21 surface regions sampled by 47  region--orbit pairs from 26 unique Mars Express orbits, and comparing HS-H simultaneously against both OMEGA and CRISM. \\
The main results of this work are summarized as follows:

\begin{itemize}

\item The wavelength-dependent correction factors derived from normalized spectra are  close to unity for bands 3--25 (0.688--0.952~$\mu$m), with a standard deviation below 1\% for both the OMEGA and CRISM comparisons, confirming that the relative spectral response of HS-H is in good agreement with both reference instruments across the operational wavelength range. Bands~1 and~2 (0.661 and 0.670~$\mu$m) show a systematic deficit of  4.7\% and 3.6\%, which may be attributed to a thermally-induced wavelength drift of the Fabry-P\'erot filter transfer functions, consistent with the temperature difference between ground calibration and in-flight operating conditions \citep{Popescu2025}.

\item The radiometric correction factors indicate that HS-H measures fluxes approximately  8\% lower than OMEGA, while showing very good agreement with CRISM ($<$1\% offset). Both offsets are practically wavelength-independent across bands 3 to 25, indicating that the radiometric discrepancy is multiplicative and does not distort the spectral shape. The mutual offset between OMEGA and CRISM ( $\sim$7.3\%) may reflect differences in their independent absolute radiometric calibrations, differences in photometric correction procedures, or residual geometric effects — the origin cannot be determined unambiguously from the present data.

\item The cross-calibration demonstrates excellent repeatability and reliability. The relative correction factors show a mean inter-image standard deviation of  0.22\% for HS-H/OMEGA and  0.20\% for HS-H/CRISM, while the absolute flux correction factors show  1.37\% and 1.70\%, respectively, across three images acquired at distances ranging from 14521 to 20924 km from Mars, with significantly different viewing geometries. These values confirm that HS-H is a stable and reliable instrument whose radiometric response does not depend significantly on distance or flyby geometry.

\item The successful application of the geometric distortion correction derived by \citet{Prodan2026} to the HS-H Mars images, enabling accurate co-registration of the science data with the reference instrument datasets, confirms the validity of the in-flight geometric calibration for planetary surface observations.

\item The photometric correction applied to the HS-H data consists of a Lambertian model based on division by the cosine of the local incidence angle. The Martian surface is known to deviate from Lambertian behavior \citep{Erard2001, Munaretto2021}, which may contribute to the observed dispersion in the absolute correction factors.

\end{itemize}

The correction factors presented in this work are subject to some limitations. The three HS-H images were acquired during a single observing session, precluding an assessment of long-term stability. The HS-H observations are not simultaneous with those of OMEGA and CRISM, and seasonal or long-term surface variations may contribute to the observed dispersion. The photometric correction applied to the HS-H data 
consists of a Lambertian model based on division by the cosine of the local incidence angle. In the absence of photometric models specifically calibrated for the analyzed regions, a quantitative correction for the geometry dependence of the absolute flux offset was not possible. At full spectral and spatial resolution, each of the 25 spectral samples of a macropixel is provided by a single sub-pixel in a single exposure, which corresponds to a signal-to-noise ratio of about 20 to 33 per spectral sample. The correction factors presented here were derived from averages of 90 to 105 macropixels, so their application to individual macropixel spectra may require spatial or spectral binning in order to reach a comparable signal-to-noise ratio. \\

The wavelength-dependent correction factors derived in this work — both normalized (Table~\ref{tab:CF_norm}) and in absolute flux (Table~\ref{tab:CF_flux}) — are directly applicable to future scientific observations of HS-H during the proximity operations at the Didymos--Dimorphos system, providing the radiometric foundation required for the compositional and spectral characterization of this binary asteroid system.

\FloatBarrier

\small{
\section*{Acknowledgements}
\textit{The authors acknowledge the ESA/Hera Flight Control and Flight Dynamics team for successfully implementing all the proposed requirements. A.F. acknowledges the University of Padova for the support through a doctoral fellowship. A.F., M.L. and F.L.F. are grateful to the Italian Space Agency (ASI) for financial support through agreement No. 2022-8-HH.0 for ESA's Hera mission. MP, JdL and GPP acknowledge financial support from ESA contract RFQ/3-18365/23/NL/GLC/my in response to Tender Action 3-18365 ``Hera Investigation Team Instrument Operations and Data Analysis Preparation''. MP acknowledges financial support from ESA/PRODEX Experiment Arrangement No. 4000151670 ``Demosaicking of Hyperspectral Observations for Solar System Small-Body Exploration''. MP acknowledges the use of computational resources and infrastructure provided by the Institute of Space Science -- INFLPR Subsidiary (Romania), as well as the technical support of Dr. Adrian Sevcenco and Dr. Ionel Stan. We thank the two referees for their careful and constructive reviews, which improved the quality of this paper.}
}

\section*{Declaration of competing interest}
The authors declare that they have no known competing financial interests or personal relationships that could have appeared to influence the work reported in this paper.

\section*{Declaration of Generative AI and AI-assisted technologies in the writing process}
During the preparation of this work the author(s) used Claude (Anthropic) in order to perform language and readability corrections. After using this tool/service, the author(s) reviewed and edited the content as needed and take(s) full responsibility for the content of the published article.

\section*{Data availability}
The OMEGA and CRISM data used in this study are publicly available through the ESA Planetary Science Archive (\url{https://psa.esa.int/psa/}) and the PDS Geosciences Node (\url{https://pds-geosciences.wustl.edu/missions/mro/crism.htm}), respectively. The HS-H data are part of the Hera mission archive and are subject to the ESA Hera data policy; access can be requested from the corresponding author. The extracted spectra and derived correction factors used in this work are available from the corresponding author upon request.


\bibliographystyle{elsarticle-harv} 
\bibliography{hsh.bib}

\FloatBarrier

\appendix

\section{Additional material}

\clearpage

\onecolumn
\begin{landscape}
\footnotesize
\setlength{\tabcolsep}{3pt}
\begin{longtable}{llccccccccccccccccccccc}

\caption{Observational geometry for all 47 OMEGA observations and the corresponding CRISM and HS-H data used in this work. Each row corresponds to one ORB--region pair. The columns report: the surface region name; the OMEGA orbit identifier (Obs.); the solar longitude of the OMEGA observation ($L_\mathrm{s}$); the latitude and longitude boundaries of the selected region (Lat$_\mathrm{min}$, Lat$_\mathrm{max}$, Lon$_\mathrm{min}$, Lon$_\mathrm{max}$); the mean phase ($\bar{\alpha}$), emission ($\bar{e}$), and incidence ($\bar{i}$) angles of OMEGA over the selected region (subscript OM); the mean phase, emission, and incidence angles of CRISM over the selected region (subscript CR); and the mean phase, emission, and incidence angles of HS-H over the selected region as measured in images HS-L0-FE-025D-04-01, HS-L0-FE-025D-05-01, and HS-L0-FE-025D-06-01 (subscripts HH4, HH5, and HH6, respectively). Dashes (---) indicate that no corresponding observation is available. All angles are given in degrees.}
\label{tab:geometry_appendix} \\

\hline\hline
Region & Obs. & $L_\mathrm{s}$ & Lat$_\mathrm{min}$ & Lat$_\mathrm{max}$ & Lon$_\mathrm{min}$ & Lon$_\mathrm{max}$ & $\bar{\alpha}_\mathrm{OM}$ & $\bar{e}_\mathrm{OM}$ & $\bar{i}_\mathrm{OM}$ & $\bar{\alpha}_\mathrm{CR}$ & $\bar{e}_\mathrm{CR}$ & $\bar{i}_\mathrm{CR}$ & $\bar{\alpha}_\mathrm{HH4}$ & $\bar{e}_\mathrm{HH4}$ & $\bar{i}_\mathrm{HH4}$ & $\bar{\alpha}_\mathrm{HH5}$ & $\bar{e}_\mathrm{HH5}$ & $\bar{i}_\mathrm{HH5}$ & $\bar{\alpha}_\mathrm{HH6}$ & $\bar{e}_\mathrm{HH6}$ & $\bar{i}_\mathrm{HH6}$ \\
 & & (°) & (°) & (°) & (°) & (°) & (°) & (°) & (°) & (°) & (°) & (°) & (°) & (°) & (°) & (°) & (°) & (°) & (°) & (°) & (°) \\
\hline
\endfirsthead

\hline\hline
Region & Obs. & $L_\mathrm{s}$ & Lat$_\mathrm{min}$ & Lat$_\mathrm{max}$ & Lon$_\mathrm{min}$ & Lon$_\mathrm{max}$ & $\bar{\alpha}_\mathrm{OM}$ & $\bar{e}_\mathrm{OM}$ & $\bar{i}_\mathrm{OM}$ & $\bar{\alpha}_\mathrm{CR}$ & $\bar{e}_\mathrm{CR}$ & $\bar{i}_\mathrm{CR}$ & $\bar{\alpha}_\mathrm{HH4}$ & $\bar{e}_\mathrm{HH4}$ & $\bar{i}_\mathrm{HH4}$ & $\bar{\alpha}_\mathrm{HH5}$ & $\bar{e}_\mathrm{HH5}$ & $\bar{i}_\mathrm{HH5}$ & $\bar{\alpha}_\mathrm{HH6}$ & $\bar{e}_\mathrm{HH6}$ & $\bar{i}_\mathrm{HH6}$ \\
 & & (°) & (°) & (°) & (°) & (°) & (°) & (°) & (°) & (°) & (°) & (°) & (°) & (°) & (°) & (°) & (°) & (°) & (°) & (°) & (°) \\
\hline
\endhead

\hline
\endfoot

Huygens & 1241\_4  & 139.5 & -15.0 & -13.0 & 53.6 & 54.7 & 39.4 & 3.5 & 40.7 & 57.5 & 0.8 & 57.8 & 20.1 & 29.2 & 44.2 & 25.7 & 26.6 & 45.2 & 34.6 & 23.3 & 46.3 \\
Huygens & 7482\_3  & 3.4 & -15.0 & -13.0 & 52.4 & 54.3 & 38.6 & 3.7 & 37.6 & 55.8 & 0.7 & 56.1 & 20.1 & 28.2 & 43.5 & 25.7 & 25.5 & 44.5 & 34.6 & 22.2 & 45.6 \\
\hline
Schiaparelli & 9850\_2  & 4.1 & -4.5 & -1.0 & 14.2 & 17.8 & 27.1 & 6.1 & 27.1 & 50.2 & 1.6 & 50.4 & 23.9 & 20.7 & 25.9 & 30.7 & 22.8 & 25.2 & 41.0 & 28.5 & 25.5 \\
Schiaparelli & 1505\_4  & 178.5 & -4.5 & -1.8 & 17.2 & 19.5 & 27.2 & 5.0 & 25.4 & 48.7 & 0.9 & 49.1 & 23.7 & 17.8 & 25.2 & 30.4 & 19.9 & 24.6 & 40.9 & 26.3 & 24.4 \\
\hline
Henry & 0558\_4  & 53.1 & 9.8 & 11.6 & 22.4 & 24.4 & 54.6 & 5.4 & 54.0 & 46.9 & 5.1 & 49.2 & 25.9 & 19.0 & 10.7 & 33.0 & 25.2 & 10.2 & --- & --- & --- \\
Henry & 1472\_3  & 173.4 & 9.8 & 11.6 & 22.8 & 24.4 & 20.6 & 4.3 & 20.4 & 45.5 & 5.8 & 48.8 & 25.8 & 18.8 & 10.5 & 33.0 & 24.9 & 10.2 & --- & --- & --- \\
Henry & 7750\_3  & 39.5 & 9.8 & 11.6 & 22.8 & 24.4 & 23.7 & 3.3 & 23.3 & 45.5 & 5.8 & 48.8 & 25.8 & 18.8 & 10.5 & 33.0 & 24.9 & 10.2 & --- & --- & --- \\
\hline
Region1 & 1472\_5  & 173.4 & -22.0 & -20.5 & 20.0 & 22.0 & 34.6 & 5.2 & 34.6 & 58.9 & 0.6 & 58.9 & 20.3 & 27.2 & 42.4 & 26.3 & 23.9 & 42.2 & 35.7 & 21.2 & 42.0 \\
Region1 & 1505\_5  & 178.5 & -22.0 & -20.0 & 20.0 & 22.0 & 32.5 & 5.7 & 31.4 & 58.6 & 0.6 & 58.6 & 20.3 & 27.0 & 42.1 & 26.3 & 23.7 & 41.9 & 35.8 & 21.0 & 41.8 \\
Region1 & 9850\_2  & 4.1 & -22.0 & -20.0 & 20.0 & 21.6 & 27.1 & 6.1 & 27.1 & 58.2 & 0.6 & 58.2 & 20.3 & 27.1 & 42.2 & 26.3 & 23.9 & 41.9 & 35.8 & 21.3 & 41.8 \\
\hline
Region2 & 1472\_4  & 173.4 & -9.8 & -7.7 & 20.5 & 21.7 & 24.8 & 4.8 & 24.7 & 53.6 & 1.1 & 53.8 & 22.5 & 16.9 & 30.2 & 29.0 & 16.3 & 29.9 & 39.2 & 20.3 & 29.7 \\
Region2 & 1505\_4  & 178.5 & -9.8 & -7.7 & 19.8 & 21.7 & 27.2 & 5.0 & 25.4 & 53.5 & 1.3 & 53.9 & 22.5 & 17.2 & 30.2 & 29.0 & 16.7 & 29.9 & 39.2 & 20.6 & 29.6 \\
Region2 & 8648\_3  & 159.7 & -9.0 & -7.7 & 20.4 & 21.7 & 29.1 & 3.1 & 28.5 & 53.5 & 1.8 & 54.2 & 22.5 & 16.8 & 29.8 & 29.1 & 16.4 & 29.5 & 39.3 & 20.5 & 29.3 \\
Region2 & 9850\_2  & 4.1 & -9.8 & -7.7 & 19.8 & 21.7 & 27.1 & 6.1 & 27.1 & 53.5 & 1.3 & 53.9 & 22.5 & 17.2 & 30.2 & 29.0 & 16.7 & 29.9 & 39.2 & 20.6 & 29.6 \\
\hline
Region3 & 1505\_3  & 178.5 & 2.0 & 4.0 & 17.8 & 19.3 & 28.6 & 4.8 & 26.4 & 50.3 & 0.6 & 50.5 & 24.8 & 18.6 & 19.6 & 31.7 & 22.7 & 18.9 & --- & --- & --- \\
Region3 & 7825\_4  & 49.2 & 2.0 & 4.0 & 17.8 & 19.3 & 29.4 & 3.5 & 29.2 & 50.3 & 0.6 & 50.5 & 24.8 & 18.6 & 19.6 & 31.7 & 22.7 & 18.9 & --- & --- & --- \\
Region3 & 8002\_0  & 71.7 & 2.0 & 4.0 & 17.8 & 18.8 & 50.7 & 4.0 & 50.2 & 50.9 & 0.6 & 51.0 & 24.8 & 19.0 & 19.7 & 31.8 & 23.0 & 19.0 & --- & --- & --- \\
Region3 & 9850\_2  & 4.1 & 2.0 & 4.0 & 17.8 & 19.3 & 27.1 & 6.1 & 27.1 & 50.3 & 0.6 & 50.5 & 24.8 & 18.6 & 19.6 & 31.7 & 22.7 & 18.9 & --- & --- & --- \\
\hline
Region4 & 1505\_3  & 178.5 & 15.7 & 16.8 & 20.3 & 22.3 & 28.6 & 4.8 & 26.4 & 49.8 & 3.8 & 50.5 & 26.9 & 25.8 & 7.2 & --- & --- & --- & --- & --- & --- \\
Region4 & 7825\_4  & 49.2 & 15.7 & 17.7 & 20.3 & 21.4 & 29.4 & 3.5 & 29.2 & 49.1 & 3.8 & 50.0 & 27.0 & 26.5 & 7.3 & --- & --- & --- & --- & --- & --- \\
Region4 & 8648\_4  & 159.7 & 15.7 & 17.7 & 20.3 & 21.0 & 37.3 & 3.3 & 36.9 & 49.6 & 2.2 & 49.1 & 27.0 & 26.7 & 7.4 & --- & --- & --- & --- & --- & --- \\
Region4 & 9850\_1  & 4.1 & 15.7 & 17.7 & 20.3 & 22.3 & 21.7 & 4.7 & 21.4 & 49.8 & 3.8 & 50.6 & 27.0 & 26.3 & 7.1 & --- & --- & --- & --- & --- & --- \\
\hline
Region5 & 1157\_5  & 128.1 & -22.0 & -20.5 & 28.9 & 29.3 & 46.7 & 2.2 & 46.8 & 62.1 & 0.5 & 62.1 & 19.8 & 23.7 & 41.8 & 25.6 & 19.0 & 41.9 & 34.8 & 13.3 & 42.1 \\
Region5 & 5418\_4  & 48.1 & -22.0 & -20.5 & 26.9 & 29.0 & 47.5 & 4.6 & 47.9 & 59.1 & 0.8 & 58.9 & 19.8 & 24.2 & 42.1 & 25.7 & 19.6 & 42.1 & 35.0 & 14.3 & 42.3 \\
\hline
Region6 & 0359\_1  & 27.7 & -15.0 & -13.0 & 36.3 & 36.7 & 44.0 & 2.6 & 44.1 & 52.1 & 0.6 & 52.5 & 20.7 & 15.2 & 35.8 & 26.6 & 9.6 & 36.3 & 36.2 & 1.0 & 36.8 \\
Region6 & 5310\_3  & 34.3 & -14.0 & -12.0 & 34.6 & 34.9 & 43.5 & 3.6 & 46.1 & 62.0 & 0.9 & 61.6 & 20.9 & 13.6 & 34.5 & 27.0 & 8.0 & 34.8 & 36.6 & 2.1 & 35.3 \\
\hline
Region7 & 1450\_3  & 170.0 & -2.7 & -0.5 & 25.6 & 28.2 & 19.9 & 4.9 & 17.5 & 51.7 & 0.6 & 51.8 & 23.4 & 7.7 & 22.1 & 30.1 & 11.0 & 22.1 & 40.6 & 19.8 & 22.3 \\
Region7 & 1472\_3  & 173.4 & -2.7 & -0.5 & 25.6 & 26.3 & 20.6 & 4.3 & 20.4 & 49.5 & 0.6 & 49.6 & 23.5 & 9.1 & 22.1 & 30.2 & 12.2 & 22.0 & 40.7 & 20.7 & 22.1 \\
Region7 & 3648\_4  & 133.2 & -2.7 & -0.5 & 25.2 & 28.2 & 35.0 & 7.7 & 34.0 & 51.6 & 0.6 & 51.7 & 23.4 & 8.0 & 22.1 & 30.1 & 11.2 & 22.1 & 40.6 & 20.0 & 22.2 \\
\hline
Region8 & 1428\_1  & 166.6 & 9.0 & 11.0 & 29.7 & 31.0 & 19.9 & 3.1 & 20.0 & 41.1 & 0.7 & 41.4 & 25.4 & 14.5 & 10.9 & 32.4 & 21.1 & 11.4 & --- & --- & --- \\
Region8 & 3648\_4  & 133.2 & 9.0 & 11.0 & 28.0 & 30.0 & 35.0 & 7.7 & 34.0 & 44.1 & 2.4 & 45.9 & 25.5 & 15.1 & 10.6 & 32.5 & 21.7 & 11.0 & --- & --- & --- \\
Region8 & 5346\_3  & 38.9 & 9.0 & 11.0 & 31.7 & 32.1 & 40.3 & 3.1 & 38.6 & 50.7 & 2.5 & 49.2 & 25.3 & 14.1 & 11.5 & 32.3 & 20.7 & 12.2 & --- & --- & --- \\
\hline
Region9 & 0426\_2  & 36.6 & -12.6 & -10.2 & 42.1 & 42.8 & 45.3 & 2.8 & 45.0 & 55.2 & 0.5 & 55.2 & 20.9 & 15.7 & 35.2 & 26.9 & 11.7 & 36.0 & 36.3 & 8.5 & 36.8 \\
Region9 & 3344\_4  & 93.9 & -13.0 & -10.3 & 43.6 & 44.8 & 39.6 & 2.9 & 38.7 & 54.2 & 0.6 & 54.4 & 20.8 & 17.4 & 36.2 & 26.7 & 13.8 & 37.0 & 36.1 & 10.6 & 37.9 \\
Region9 & 5443\_4  & 51.3 & -12.0 & -10.3 & 43.6 & 44.8 & 43.1 & 3.5 & 41.9 & 53.7 & 0.6 & 54.0 & 20.9 & 16.9 & 35.7 & 26.8 & 13.4 & 36.5 & 36.2 & 10.7 & 37.4 \\
\hline
Region13 & 0554\_4  & 52.6 & 7.0 & 9.6 & 53.3 & 55.3 & 53.8 & 5.2 & 53.1 & 50.5 & 0.6 & 50.6 & 24.3 & 27.7 & 28.8 & 30.8 & 30.9 & 30.3 & 40.8 & 36.6 & 31.8 \\
Region13 & 7482\_1  & 3.4 & 7.0 & 9.6 & 54.0 & 54.3 & 34.3 & 4.4 & 31.5 & 52.6 & 0.6 & 52.6 & 24.3 & 27.3 & 28.5 & 30.8 & 30.5 & 30.0 & 40.8 & 36.3 & 31.5 \\
\hline
Region17 & 0426\_2  & 36.6 & -8.0 & -6.0 & 41.2 & 42.5 & 45.3 & 2.8 & 45.0 & 52.1 & 0.6 & 52.2 & 21.7 & 11.8 & 30.8 & 27.9 & 9.5 & 31.6 & 37.7 & 11.5 & 32.5 \\
\hline
Region19 & 0359\_2  & 27.7 & -4.0 & -1.9 & 36.0 & 36.9 & 30.4 & 2.7 & 30.0 & 48.1 & 0.6 & 48.1 & 22.7 & 3.9 & 24.8 & 29.2 & 5.7 & 25.4 & 39.3 & 14.3 & 26.1 \\
\hline
Region21 & 0359\_3  & 27.7 & 16.4 & 18.4 & 33.0 & 34.5 & 26.9 & 4.6 & 26.7 & 47.0 & 3.5 & 49.6 & 26.6 & 22.9 & 6.8 & --- & --- & --- & --- & --- & --- \\
\hline
Region22 & 0554\_4  & 52.6 & 13.0 & 16.0 & 51.0 & 53.0 & 53.8 & 5.2 & 53.1 & 49.6 & 0.6 & 49.7 & 25.5 & 29.2 & 23.9 & 32.2 & 33.8 & 25.5 & --- & --- & --- \\
\hline
Region23 & 0294\_3  & 18.1 & 16.0 & 16.6 & 40.7 & 42.5 & 17.0 & 2.8 & 16.9 & 48.4 & 1.0 & 48.0 & 26.1 & 23.8 & 14.1 & 33.1 & 29.9 & 15.6 & --- & --- & --- \\
Region23 & 0426\_4  & 36.6 & 16.5 & 19.6 & 38.6 & 42.0 & 35.1 & 4.7 & 34.6 & 49.3 & 0.9 & 49.0 & 26.4 & 25.1 & 12.5 & 33.2 & 30.2 & 15.2 & --- & --- & --- \\
\hline
Region24 & 3721\_3  & 143.3 & 22.8 & 26.0 & 50.2 & 51.4 & 39.0 & 5.8 & 43.8 & 56.9 & 4.4 & 56.0 & 27.2 & 37.3 & 22.4 & --- & --- & --- & --- & --- & --- \\
\hline
Region26 & 5483\_3  & 56.3 & 5.0 & 8.0 & 3.2 & 4.6 & 19.6 & 3.4 & 21.8 & 47.0 & 0.8 & 47.3 & 26.2 & 36.1 & 26.9 & 33.3 & 39.8 & 25.9 & --- & --- & --- \\
\hline
Region30 & 1479\_3  & 174.5 & -2.4 & 1.0 & 54.0 & 57.5 & 21.8 & 4.1 & 21.6 & 52.7 & 0.8 & 52.8 & 22.6 & 26.9 & 35.2 & 28.7 & 27.7 & 36.6 & 38.3 & 30.2 & 38.0 \\
Region30 & 7482\_2  & 3.4 & -2.4 & 1.0 & 53.3 & 54.4 & 31.1 & 3.6 & 29.2 & 52.1 & 0.5 & 52.1 & 22.6 & 24.9 & 33.7 & 28.8 & 25.8 & 35.1 & 38.4 & 28.6 & 36.4 \\
\hline
\end{longtable}
\end{landscape}

\begin{figure}
\centering
\caption{
Median HS-H spectra extracted from the selected Martian regions used in the analysis, from image HS-L0-FE-025D-04-01. Each panel shows one spectrum associated with a specific region and OMEGA observation. The x-axis reports the wavelength in microns, while the y-axis reports the reflectance factor. For clarity, each spectrum is shown with an independent y-axis scale. The region name, the corresponding OMEGA orbit identifier, and the number of macropixels used for the spectral extraction are reported inside each panel. Only the spectra retained after the quality selection are shown, for a total of 47 spectra.
}
\label{fig:appendix2}
\includegraphics[width=1.0\textwidth]{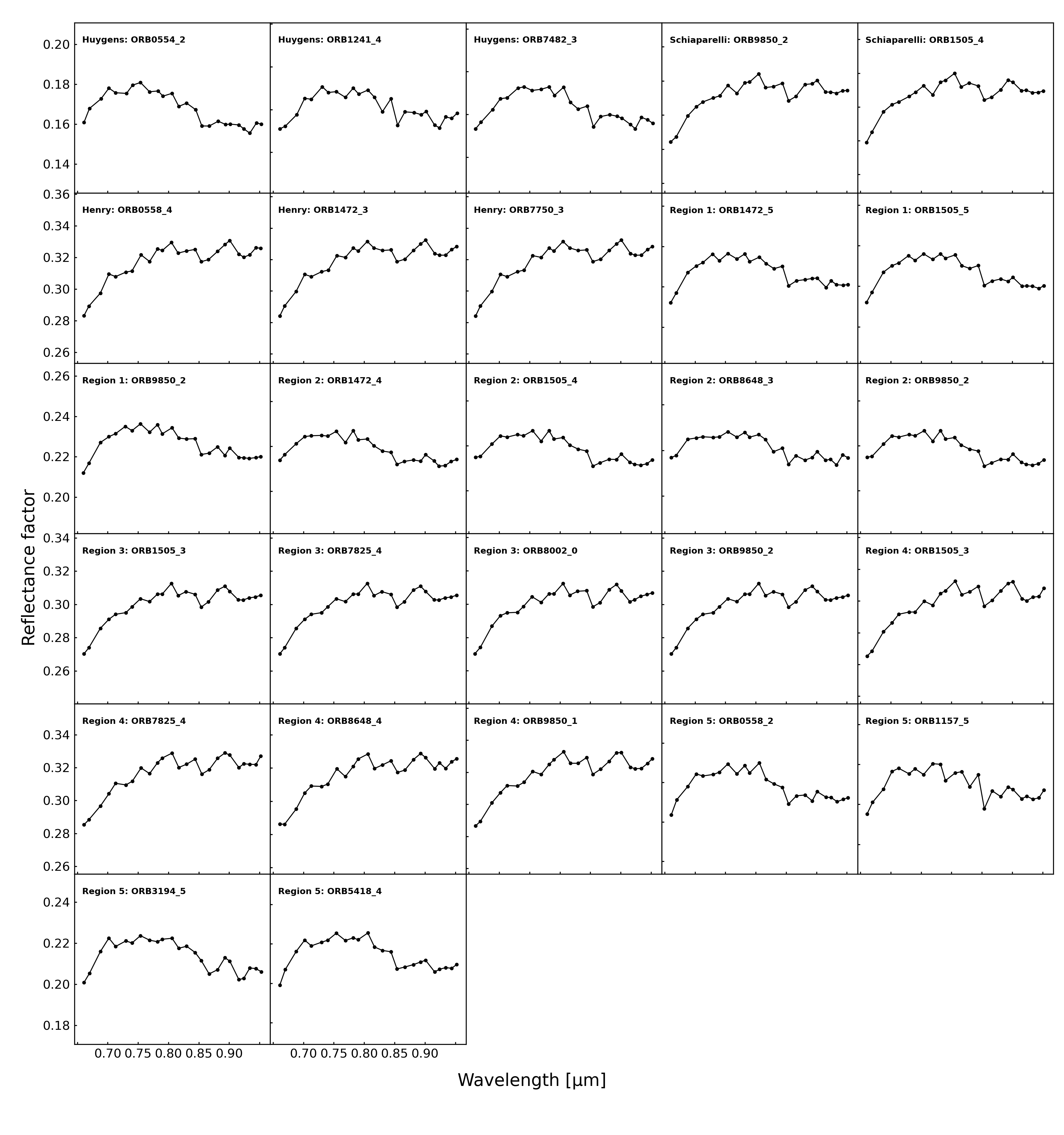}
\end{figure}

\begin{figure}\ContinuedFloat
\centering
\includegraphics[width=1.0\textwidth]{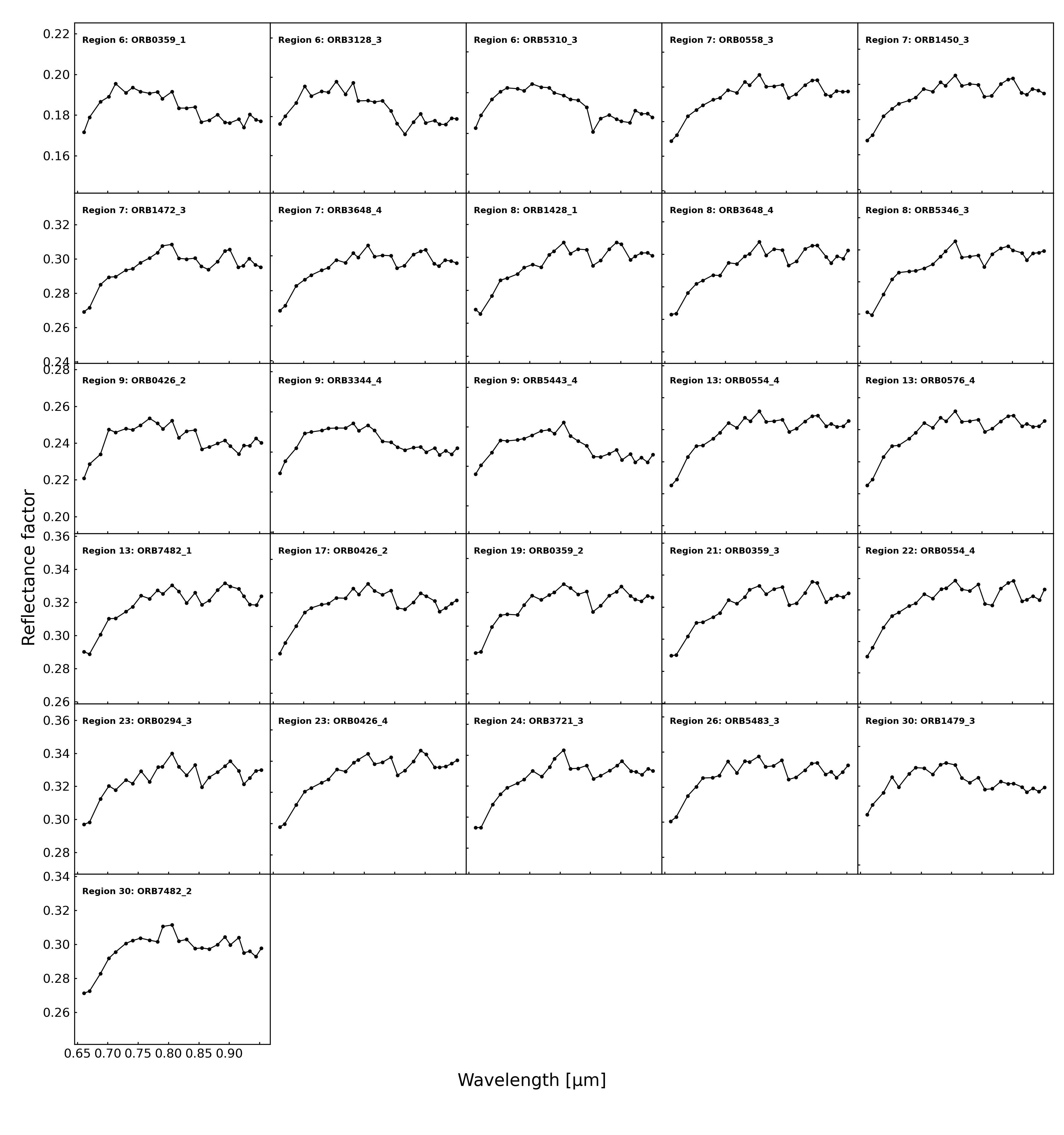}
\end{figure}

\twocolumn

\clearpage

\end{document}